\documentclass[a4paper,preprintnumbers,showpacs,onecolumn,superscriptaddress,nofootinbib,amsmath,amssymb,notitlepage]{revtex4-2}

\usepackage{mathtools,leftindex,tensor,mhchem}
\usepackage{float}
\usepackage{booktabs}
\usepackage{enumitem}
\usepackage{times}
\usepackage{dcolumn}
\usepackage{mathrsfs}
\usepackage{comment}
\usepackage{multirow}
\usepackage{hyperref}
\usepackage{amsmath,accents}
\usepackage{mathtools}
\usepackage{bbm}
\usepackage{bm}
\usepackage{amsfonts}
\usepackage{amssymb}
\usepackage{mathrsfs}
\usepackage[scr=rsfs,cal=boondox]{mathalfa}
\usepackage{wasysym}
\usepackage{graphicx}
\usepackage{filecontents}
\usepackage[dvipsnames]{xcolor}
\usepackage{bbold}
\usepackage[scr=rsfs,cal=boondox]{mathalfa}

\usepackage{stackengine}
\stackMath
\newcommand\tsup[2][2]{%
 \def\useanchorwidth{T}%
  \ifnum#1>1%
    \stackon[-1.3ex]{\tsup[\numexpr#1-1\relax]{#2}}{\mathchar"307E}%
  \else%
    \stackon[-1ex]{#2}{\mathchar"307E}%
  \fi%
}
\hypersetup{
    bookmarks=true,         % show bookmarks bar?
    pdftoolbar=true,        % show Acrobats toolbar?
    pdfmenubar=true,        % show Acrobats menu?
    pdffitwindow=true,     %window fit to page when opened
    pdfstartview={FitH},     %fits the width of the page to the window
    pdftitle={My title},     %title
    pdfauthor={author},      %author
    pdfsubject={Subject},    %subject of the document
    pdfcreator={Creator},    %creator of the document
    pdfproducer={Producer},  %producer of the document
    pdfkeywords={keyword1} %{key2} {key3}, % list of keywords
    pdfnewwindow=true,       %links in new window
    colorlinks=true ,        %false: boxed links; true: colored links
    linkcolor=Blue  ,        %color of internal links (change box color with linkbordercolor)
    citecolor=Blue,      %color of links to bibliography
    filecolor=Blue,       %color of file links
    urlcolor=Blue            %color of external links
}

\newcommand{\ed}{\mathrm{d}}

\newcommand{\mL}{\mathcal{L}}
\newcommand{\mc}{\mathcal{c}}

\newcommand{\R}{\mathcal{R}}

\newcommand{\ba}{\bar{a}}

\newcommand{\oalpha}[1]{\accentset{\circ}{\alpha}}
\newcommand{\obf}[1]{\accentset{\circ}{\mathbf{f}}}
\newcommand{\boR}[1]{\accentset{\circ}{\mathbf{R}}}
\newcommand{\obF}[1]{\accentset{\circ}{\mathbf{F}}}
\newcommand{\obPi}[1]{\accentset{\circ}{\mathbf{\Pi}}}

\usepackage{scalerel}
\usepackage{tikz}
\usetikzlibrary{svg.path}

\definecolor{orcidlogocol}{HTML}{A6CE39}
\tikzset{
  orcidlogo/.pic={
    \fill[orcidlogocol] svg{M256,128c0,70.7-57.3,128-128,128C57.3,256,0,198.7,0,128C0,57.3,57.3,0,128,0C198.7,0,256,57.3,256,128z};
    \fill[white] svg{M86.3,186.2H70.9V79.1h15.4v48.4V186.2z}
                 svg{M108.9,79.1h41.6c39.6,0,57,28.3,57,53.6c0,27.5-21.5,53.6-56.8,53.6h-41.8V79.1z M124.3,172.4h24.5c34.9,0,42.9-26.5,42.9-39.7c0-21.5-13.7-39.7-43.7-39.7h-23.7V172.4z}
                 svg{M88.7,56.8c0,5.5-4.5,10.1-10.1,10.1c-5.6,0-10.1-4.6-10.1-10.1c0-5.6,4.5-10.1,10.1-10.1C84.2,46.7,88.7,51.3,88.7,56.8z};
  }
}

\newcommand\orcidicon[1]{\href{https://orcid.org/#1}{\mbox{\scalerel*{
\begin{tikzpicture}[yscale=-1,transform shape]
\pic{orcidlogo};
\end{tikzpicture}
}{|}}}}
\begin{document}

%%%%%%
\title{Effect of primary scalar hair on black hole's strong lensing in Beyond Horndeski gravity}

\author{Mohsen Fathi\orcidicon{0000-0002-1602-0722}}
\email{mohsen.fathi@ucentral.cl}
\affiliation{Centro de Investigaci\'{o}n en Ciencias del Espacio y F\'{i}sica Te\'{o}rica, Universidad Central de Chile, La Serena 1710164, Chile}

%%%%%%%%%%%abstract

\begin{abstract}

We study the gravitational lensing effects of a static, asymptotically flat black hole with primary scalar hair in Beyond Horndeski gravity, focusing on the strong lensing regime. Recently, in Ref. \cite{erices_thermodynamic_2025}, we placed constraints on the scalar hair parameter by analyzing its thermodynamic stability and the black hole shadow. In this work, we further investigate the strong lensing properties of the black hole, which form the basis of shadow formation, and employ observational data from the Event Horizon Telescope to derive more precise constraints on the scalar hair parameter. Additionally, we compute the shape and position of different lensed images of a thin accretion disk and argue that the observed black hole shadow corresponds to the secondary image of the emitting disk. Using this interpretation, we perform a new set of constraints on the scalar hair. Furthermore, we discuss why higher-order images are not suitable for astrophysical constraints, highlighting the limitations posed by their faintness and observational challenges. Finally, we find that models satisfying these constraints exhibit local instabilities.

\bigskip

{\noindent{\textit{keywords}}: Black holes, Beyond Horndeski gravity, scalar hair, strong lensing, photon rings
}\\

\end{abstract}

\maketitle
%%%%%%%%%%%%%%%%%%%%%%%%%%%%%%%%%%
\section{Introduction}

Black holes represent one of the most remarkable predictions of General Relativity. Recent advancements in observational astronomy have provided groundbreaking confirmation of their existence. The emergence of gravitational-wave astronomy was marked by the first direct detection of gravitational waves, followed by the observation of a binary black hole merger~\cite{Abbott1}. Additionally, the concurrent detection of gravitational waves and gamma rays from a binary neutron star merger offered compelling evidence for the luminal speed of gravitational waves~\cite{Abbott2,Abbott3}. 

Moreover, the Event Horizon Telescope (EHT) collaboration succeeded in capturing the first direct images of supermassive black holes, initially for M87*~\cite{eht1,Akiyama:2019,the_event_horizon_telescope_collaboration_first_2019,the_event_horizon_telescope_collaboration_first_2019-1} and later for Sgr A* at the center of the Milky Way~\cite{Akiyama:2022,event_horizon_telescope_collaboration_first_2022-1}. These findings were further enhanced through polarization imaging of the emission ring~\cite{eht3,ehtsa2}, reinforcing general relativity’s status as the most successful gravitational theory for over a century.

Despite its triumphs, general relativity remains incomplete, as it fails to reconcile gravitational interactions at both the microscopic and cosmic scales. On small scales, it is experimentally inaccessible between the Planck length and the micron scale. On larger scales, unresolved issues such as the cosmological constant problem~\cite{ccproblem} and the Hubble tension~\cite{tension2,tension1,tension4,tension3,tension5} necessitate either substantial amounts of dark matter and dark energy or modifications to the theory itself.

To address these limitations, several alternative gravity theories have been proposed, including String Theory~\cite{Zw}, Emergent Gravity~\cite{verlinde}, Asymptotic Safety~\cite{reuter}, and Loop Quantum Gravity~\cite{LQG}. These frameworks aim to resolve singularity issues~\cite{singularity1,singularity2} and offer a microscopic foundation for the thermodynamic nature of gravity~\cite{Bekenstein01,H1,H2,ashtekar,strominger}. Among these alternatives, scalar-tensor theories have attracted significant interest due to their incorporation of additional scalar degrees of freedom.

Horndeski theory represents the most general class of scalar-tensor theories that maintain second-order field equations while avoiding Ostrogradski instabilities~\cite{Horndeski:1974wa}. Later advancements demonstrated that scalar-tensor theories could accommodate higher-order field equations without instability by utilizing degenerate Lagrangians. This led to the development of Beyond Horndeski and Degenerate Higher-Order Scalar–Tensor theories~\cite{Crisostomi:2016tcp,Gleyzes:2014dya,Kobayashi:2019hrl,Langlois:2015cwa,Langlois:2017mdk,Langlois:2018dxi,baake_spinning_2020,babichev_regular_2020,bravo-gaete_shear_2022,bravo-gaete_planar_2022}, and even rigorous constraints have been proposed~\cite{saltas_searching_2022}.

Obtaining black hole solutions in these modified theories is challenging due to extended no-hair theorems, which constrain the existence of non-trivial scalar field configurations~\cite{SotiriouBHs,Hui,MaselliBHs,guajardo_primary_2025}. The no-hair conjecture posits that a black hole is entirely described by its mass $M$, angular momentum $J$, and electric charge $Q$, measurable at asymptotic infinity~\cite{PhysRevLett.11.237}. A black hole with an additional global charge is said to possess primary hair, whereas secondary hair implies that the metric remains fully determined by $M$, $J$, and $Q$ despite the presence of additional fields. Both cases are often categorized under the broader concept of ``hairy black holes''~\cite{Bekenstein:1971hc,Teitelboim:1972ps,Chrusciel:2012jk,Mazur:2000pn}.

Constructing black hole solutions with primary scalar hair is particularly complex. Recently, the first such solution was identified within a shift-symmetric subclass of Beyond Horndeski theories~\cite{Bakopoulos:2023fmv}, followed by further generalizations~\cite{Baake:2023zsq,Bakopoulos:2023sdm}. These solutions introduce a conserved scalar charge, significantly affecting the black hole's observational signatures, making them valuable candidates for testing alternative gravity theories via strong lensing and shadow observations.

Ref.~\cite{erices_thermodynamic_2025} recently examined the thermodynamic stability and shadow constraints associated with this black hole solution, highlighting how the sign of the scalar hair parameter influences the shadow size. In this study, we extend this analysis by investigating the strong gravitational lensing properties of these black holes and examining the impact of scalar hair on lensing observables.

The study of gravitational lensing dates back to the early days of general relativity, with Darwin being the first to analyze the Schwarzschild black hole (SBH) and derive exact expressions for the light deflection angle, demonstrating its logarithmic divergence near the photon sphere~\cite{Darwin_gravity_1959}. Black hole lensing remains one of the most effective tools for probing strong-field gravity. Virbhadra and Ellis developed the lens equation for an SBH in the strong-field regime, predicting the formation of two infinite sequences of faint relativistic images on either side of the black hole~\cite{Virbhadra:2000}. Consequently, the study of strong gravitational lensing has garnered significant interest.

Building on the Virbhadra-Ellis lens equation~\cite{claudel_geometry_2001,Virbhadra:2002}, Bozza extended analytical lensing methods to a wide class of static and spherically symmetric spacetimes, demonstrating that logarithmic divergence of the deflection angle at the photon sphere is a common feature in such backgrounds. He also identified the primary and secondary images and reported the existence of two infinite sets of relativistic images~\cite{Bozza:2002}. These findings were later extended by considering finite observer and source positions~\cite{bozza_strong_2007}. In cases where the black hole is illuminated by a thin accretion disk, Tsupko applied Bozza's method to determine the shape of higher-order images using both analytical and numerical techniques~\cite{tsupko_shape_2022}. More recently, Ref.~\cite{aratore_constraining_2024} demonstrated that the spacing between higher-order images ({termed as photon rings in that reference}) can reveal key properties of the underlying spacetime.
{It is, however, important to clarify the terminology used for the various light rings forming around the black hole shadow. Following Ref. \cite{Gralla:2019}, images with order $n \geq 2$ are termed photon rings, while the $n=1$ image is known as the lensing ring or secondary image. In this work, the tertiary image corresponds to the $n=2$ photon ring in that reference. Notably, the general relativistic magnetohydrodynamics (GRMHD) simulations predominantly observe the lensing ring (secondary image), which appears as the photon ring in those simulations. The tertiary image, being thin and closely attached to the inner shadow, remains unresolved in these observations.}

Strong gravitational lensing remains a powerful tool for probing black hole spacetimes in both general relativity and alternative theories of gravity. Numerous studies have explored this phenomenon in various contexts \cite{bhadra_gravitational_2003,eiroa_braneworld_2005,whisker_strong_2005,eiroa_braneworld_2005-1,sarkar_strong_2006,chen_strong_2009,eiroa_gravitational_2011,li_strong_2015,chakraborty_strong_2017,javed_effect_2019,shaikh_analytical_2019,ovgun_weak_2019,panpanich_particle_2019,bronnikov_gravitational_2019,shaikh_novel_2019,kumar_gravitational_2020,islam_gravitational_2020,lu_gravitational_2021,babar_particle_2021,narzilloev_motion_2021,tsukamoto_gravitational_2021,soares_gravitational_2023,soares_holonomy_2023,vachher_probing_2024,turakhonov_observational_2024,soares_topologically_2024}.

Motivated by the considerations above regarding the choice of the gravitational theory, we structure this paper as follows: In Sect. \ref{sec:BH_solution}, we provide a concise review of the theory under investigation, introducing the static, spherically symmetric, asymptotically flat black hole solution proposed within this framework and discussing its key properties. In Sect. \ref{sec:deflection}, we initiate our analysis by formulating the deflection angle of light in the strong lensing regime, considering both the source and the observer at finite distances from the black hole. The relevant integrals are then evaluated numerically, and we examine how the deflection angle varies with the impact parameter of the passing light rays, as perceived by the observer. Sect. \ref{sec:lensEq} is dedicated to the lens equation governing relativistic images of different orders. We utilize this equation to explore the response of RERs to variations in the black hole's scalar hair. Additionally, we introduce three key observables in strong gravitational lensing and employ the angular diameter of relativistic images as a tool to assess the black hole's properties in light of the EHT data for M87* and Sgr A*. In Sect. \ref{sec:accImages}, we adopt an alternative approach to constraining the black hole parameters by considering a thin accretion disk as the source of illumination. We derive analytical expressions for the shapes of images of different orders and establish a connection between the observed black hole shadow and the primary image of the accretion disk. Using this framework, we impose new constraints on the black hole's scalar hair. Furthermore, we demonstrate why higher-order images play a less significant role in astrophysical observational tests based on EHT data. Finally, we summarize our findings in Sect. \ref{sec:conclusions}.

Throughout this study, we employ geometric units where $ G=\hbar=\ell_p=c=1$, adopt the sign convention $(-,+,+,+)$, and use primes to denote derivatives with respect to the radial coordinate.

%%%%%%%%%%%%%%%%%%%%%%%%%%%%%%%%%%%sect.1
\section{The black hole solution with primary scalar hair}\label{sec:BH_solution}

This section outlines the fundamental properties of the static, spherically symmetric black hole solution introduced in Ref.~\cite{Bakopoulos:2023fmv}. In the framework of Beyond Horndeski theories~\cite{Gleyzes:2015}, where the shift symmetry $\tilde{\chi}\rightarrow\tilde{\chi}+\text{const.}$ and parity symmetry $\tilde{\chi}\rightarrow{-\tilde{\chi}}$ are preserved, the theory is characterized by three arbitrary functions, $G_2$, $G_4$, and $F_4$, which are dependent on the kinetic term of the scalar field, given by $X=-\frac{1}{2}\tilde{\chi}_{,\mu}\tilde{\chi}^{,\mu}$. The corresponding action takes the form
\begin{equation}
I\left[g_{\mu\nu},\tilde{\chi}\right]=\frac{1}{16\pi%\kappa
}\int\ed^4x\,\sqrt{-g}\Biggl\{
G_2(X)+G_4(X)R+G_{4 X}\left[(\square \tilde{\chi})^2-\tilde{\chi}_{\mu \nu} \tilde{\chi}^{\mu \nu}\right]+F_4(X) \epsilon^{\mu \nu \rho \sigma} \epsilon^{\alpha \beta \gamma}{ }_\sigma \tilde{\chi}_\mu \tilde{\chi}_\alpha \tilde{\chi}_{\nu \beta} \tilde{\chi}_{\rho \gamma}
\Biggr\},
    \label{eq:action}
\end{equation}
where the notations $\tilde{\chi}_\mu=\partial_\mu \tilde{\chi}$, $\tilde{\chi}_{\mu \nu}=\nabla_\mu \partial_\nu \tilde{\chi}$, and subscripts denoting derivatives with respect to $X$ are used for brevity. The Horndeski functionals that support the existence of a black hole with scalar hair are given by
\begin{equation}
G_2=-\frac{8\eta}{3\lambda^2}X^2,\qquad G_4=1-\frac{4\eta}{3}X^2,\qquad F_4=\eta,
    \label{eq:functionals}
\end{equation}
where $\lambda$ has the dimension of length, while $\eta$ has the dimension of (length)$^4$. The scalar field follows the functional form
\begin{equation}
\tilde{\chi} = q t +\zeta(r),
    \label{eq:varphi}
\end{equation}
where $q$ is a dimensionful parameter with units of (length)$^{-1}$, playing the role of the primary hair.

The static, spherically symmetric spacetime describing a black hole with primary hair (hereafter referred to as the PHBH) is given by the metric
\begin{equation}
\ed s^2=-f(r)\ed t^2+\frac{\ed r^2}{f(r)}+r^2\ed\theta^2+r^2\sin^2\theta\ed\phi^2,
    \label{eq:metr0}
\end{equation}
where the metric function takes the form
\begin{equation}
f(r) = 1-\frac{2M}{r}+\mc\left[\frac{\pi/2-\arctan(r/\lambda)}{r/\lambda}+\frac{1}{1+(r/\lambda)^2}\right],
    \label{eq:lapse_0}
\end{equation}
with $\mathcal{c}=\eta q^4$ being a dimensionless parameter. The function $\zeta(r)$ is determined through the differential equation
\begin{equation}
\zeta'(r)=\pm\sqrt{\frac{q^2}{f(r)^2}\left[1-\frac{f(r)}{1+(r/\lambda)^2}\right]}.
    \label{eq:psi}
\end{equation}
Consequently, the kinetic term of the scalar field is expressed as
\begin{equation}
X=\frac{q^2/2}{1+(r/\lambda)^2}.
    \label{eq:X}
\end{equation}
Since the theory is invariant under the transformation $\lambda \rightarrow -\lambda$, we may, without loss of generality, restrict to $\lambda > 0$. Setting $M = \mc = 0$ yields the Minkowski spacetime, while the case $M = 0$ with $\mc \neq 0$ corresponds to a nontrivial geometry. {For $\mc < 0$, the solution describes a black hole, whereas $\mc > 0$ -- for sufficiently large values, as discussed below -- leads to a naked singularity. }

In the metric function \eqref{eq:lapse_0}, the constants $M$ and $q$ correspond to the Arnowitt-Deser-Misner (ADM) mass~\cite{ADM:1961} and the primary scalar hair, respectively. Observing Eq.~\eqref{eq:X}, one can see that in the absence of scalar hair ($q=0$), the modifications beyond general relativity vanish, recovering the Schwarzschild solution from Eq.~\eqref{eq:lapse_0}. At large distances, the asymptotic expansion of the lapse function \eqref{eq:lapse_0} reads
\begin{equation}
f(r) = 1-\frac{2M}{r}+2\lambda^2\frac{\mc}{r^2}+\mathcal{O}(r^{-4}),
    \label{eq:lapse_1}
\end{equation}
which corresponds to the asymptotic scalar fields $\tilde{\chi}=q v$ and $\tilde{\chi}=q u$ for the positive and negative signs in Eq.~\eqref{eq:psi}, respectively, where $u$ and $v$ denote the retarded and advanced null coordinates. The asymptotic behavior in Eq.~\eqref{eq:lapse_1} suggests that, for a positive ADM mass $M$, the solution remains asymptotically flat and closely resembles the Reissner-Nordström metric in general relativity, with $\mc$ playing a role analogous to the electric charge of a black hole.

\begin{figure}[t]
\centering
    \includegraphics[width=7cm]{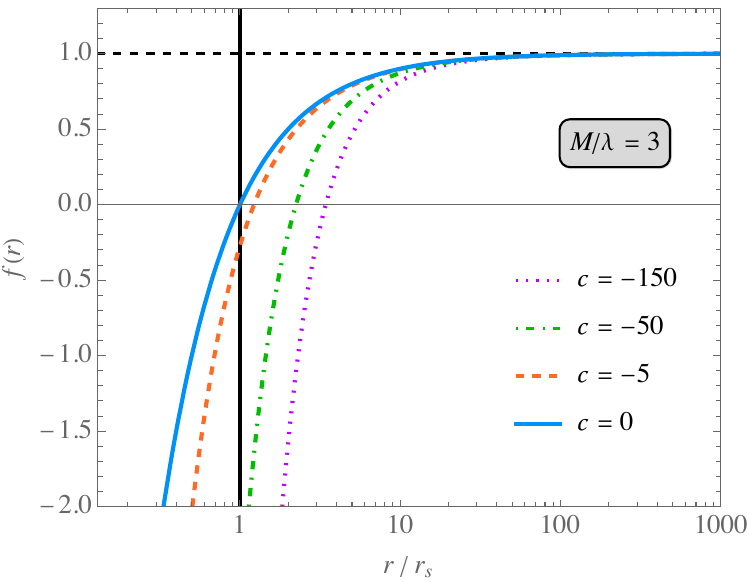} (a)\quad
    \includegraphics[width=7cm]{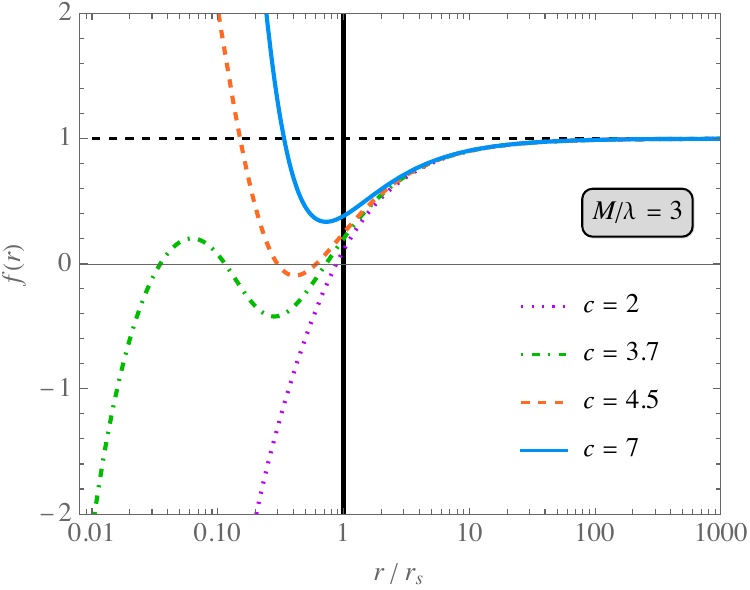} (b)
    \caption{The behavior of the lapse function $f(r)$ for $\mc\leq0$ in panel (a) and $\mc>0$ in panel (b), plotted for different values of $\mc$. While the case of $\mc<0$ corresponds to a single-horizon black hole, the case of $\mc>0$ exhibits a variety of possibilities, ranging from multiple-horizon black holes to naked singularities. The horizontal axis is logarithmic in both panels, and the unit of length along the axes is set as the Schwarzschild radius $r_s$.}
    \label{fig:f(r)}
\end{figure}

The sign of the parameter $\mc$ is determined by that of $\eta$, which significantly influences the characteristics of the solution. Specifically, for a fixed $\mc$, any alteration in $\eta$ must be compensated by a corresponding change in the scalar hair parameter $q$. Moreover, $\eta$ plays a pivotal role in defining the causal structure of the black hole, as the number of event horizons strongly depends on this parameter. In the near-origin limit ($r\rightarrow0$), the lapse function \eqref{eq:lapse_0} behaves as~\cite{Bakopoulos:2024}
\begin{equation}
f(r) = 1-\frac{2M-\pi\mc\lambda/2}{r}-\frac{2\mc r^2}{3\lambda^2}+\mathcal{O}(r^4).
    \label{eq:lapse_2}
\end{equation}
In Fig. \ref{fig:f(r)}, we present the radial profile of the lapse function \eqref{eq:lapse_0} for various values of the coefficient $\mc$, considering both negative and positive values of $\mc$.  
The SBH is recovered for $\mc=0$, with the Schwarzschild radius given by $r_s=2M$. As observed in Fig. \ref{fig:f(r)}(a), for $\mc<0$, the lapse function tends to $f(r) \to -\infty$ as the singularity at $r=0$ is approached. In this case, the black hole possesses a single event horizon, with $r_+>r_s$, indicating that black holes with negative primary hair are larger than the SBH. For $\mc>0$, the theory allows for various scenarios, from black holes with multiple horizons to naked singularities. 

As inferred from Eq. \eqref{eq:lapse_2} and depicted in Fig. \ref{fig:f(r)}(b), the nature of the solution depends critically on the ratio $M/\lambda$ compared to $\pi\mc/4$, which determines a critical value of $\mc$ given by $\mc_*=4 M/(\pi\lambda)$. When $\mc<\mc_*$, we find that $f(r) \to -\infty$ as the singularity is approached. In this regime, the event horizon $r_+$ lies within $r_s$, meaning that the black hole is more compact than its general relativistic counterpart with $\mc=0$. As $\mc$ increases, the event horizon shrinks until the emergence of three horizons, provided that $M/\lambda>1+\pi/4$ \cite{Bakopoulos:2024}. For sufficiently large primary hair, $\mc>\mc_*$, the lapse function instead diverges positively, $f(r) \to +\infty$, near the singularity. As $\mc$ increases further, the conditions for horizon formation become increasingly unfavorable. Initially, the system transitions through a two-horizon black hole phase, followed by an extremal black hole (EBH), and ultimately, for sufficiently large $\mc$, a naked singularity emerges.  Note also that, the theory is governed by the parameters $\lambda$ and $\eta$. However, as evident from the metric function in Eq. \eqref{eq:lapse_0}, the relevant dimensionless quantities for the analyses in the following sections are $\mc$ and the ratio $M/\lambda$.

%%%%%%%%%%%%%%%%%%%%%
\section{Strong gravitational lensing by the PHBH}\label{sec:deflection}

To analyze gravitational lensing effects around a black hole, it is crucial to examine the behavior of null geodesics in its vicinity. Essentially, as light rays travel closer to the black hole, the lensing effects become increasingly pronounced. Following the methodology outlined in Refs.~\cite{Bozza:2001,Bozza:2002,bozza_strong_2007} (also see Ref.~\cite{bozza_gravitational_2010}), we consider a general static and spherically symmetric spacetime with the line element
\begin{equation}
    \ed s^2=-A(r) \ed t^2+B(r) \ed r^2+D(r)\left(\ed\theta^2+\sin^2\theta\ed\phi^2\right).
    \label{eq:metric_0}
\end{equation}
By comparing this metric with the PHBH spacetime described in Eqs.~\eqref{eq:metr0} and \eqref{eq:lapse_0}, one finds the identifications $A(r)=f(r)$, $B(r)=f^{-1}(r)$, and $D(r)=r^2$. Consequently, in the asymptotic limit, we obtain $A(r)\to1$, $B(r)\to1$, and $D(r)/r^2\to1$ as $r\to\infty$. 

\begin{figure}[htp]
    \centering
    \includegraphics[width=9cm]{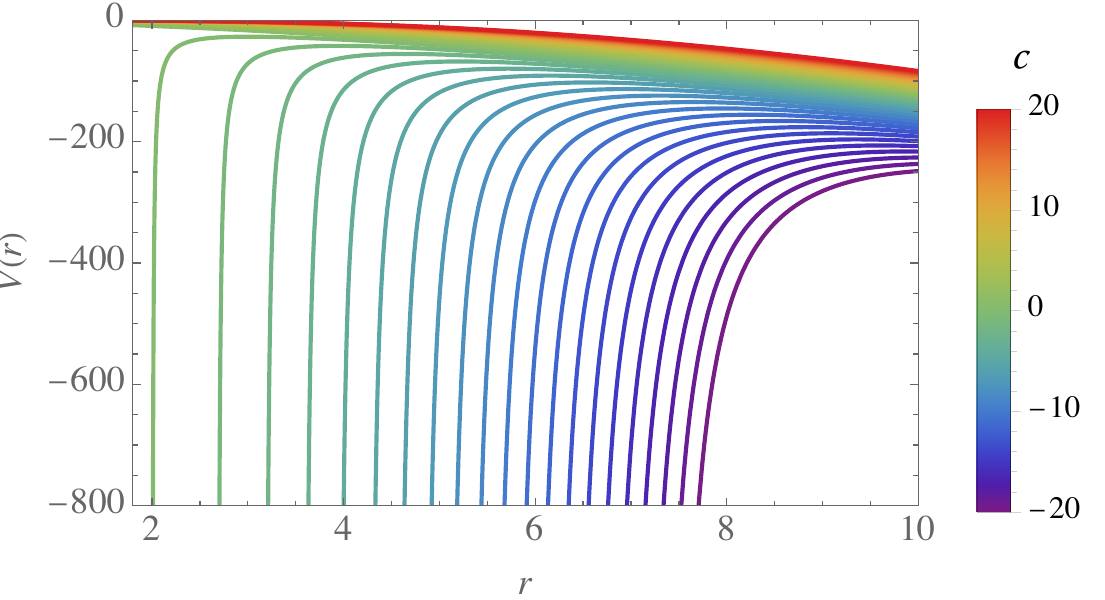}
    \caption{The radial profile of the potential $V(r)$ for the PHBH, shown for $M/\lambda=1$ and different values of the $\mc$-parameter.}
    \label{fig:V(r)}
\end{figure}

We also assume that the spacetime given by Eq.~\eqref{eq:metric_0} accommodates at least one photon sphere, where null geodesics trace spherical orbits of constant radius. The condition determining such orbits is given by $V'(r_p)=0$~\cite{atkinson_light_1965,claudel_geometry_2001,virbhadra_gravitational_2002,Bozza:2002} (also see the review in Ref.~\cite{perlick_calculating_2022}), where
\begin{equation}
    V(r)=-\frac{D(r)}{A(r)}.
    \label{eq:V(r)}
\end{equation}
Depending on whether the extremum of $V(r)$ corresponds to a maximum or minimum, the resulting orbits are categorized as unstable or stable, respectively. Given that the spacetime is asymptotically flat, we find $V(r)\to-\infty$ as $r\to\infty$. This implies that the extremum $r_m$ of $V(r)$ represents a maximum, making the orbits in the region $r_m<r<\infty$ unstable. Consequently, for all other orbits, the effective potential satisfies $V(r)<V(r_p)$. In Fig.~\ref{fig:V(r)}, we illustrate the behavior of the potential for various values of the $\mc$-parameter.
As evident from the figure, the potential exhibits a local maximum for each fixed value of $\mc$, corresponding to $r_m=r_p$, and declines for $r>r_p$. As light rays approach the photon sphere, their bending angle increases progressively. Consequently, before reaching an observer, they may complete multiple revolutions around the black hole, producing an infinite sequence of images of a single light source. In the subsequent sections and following Refs.~\cite{bisnovatyi-kogan_analytical_2022, tsupko_shape_2022, aratore_constraining_2024}, we denote these images by integer indices $n=0,1,2,3,\dots$, known as "orders," with each order corresponding to a crossing of the observer’s line of sight.

For spherically symmetric spacetimes, light rays can be assumed to propagate within the equatorial plane, defined by the polar coordinates $(r,\phi)$. The total change in the azimuthal angle $\phi$ for a light ray emitted from a source at $r_S$ and detected by an observer at $r_O$ (both measured from the black hole) is given by~\cite{bozza_strong_2007}
\begin{equation}
    \Delta\phi=\int_R^{r_S}b\sqrt{\frac{B(r)}{D(r)\R(r,b)}}\,\ed r+\int_R^{r_O}b\sqrt{\frac{B(r)}{D(r)\R(r,b)}}\,\ed r,
    \label{eq:Deltaphi_0}
\end{equation}
where we adopt the notation of Ref.~\cite{aratore_constraining_2024}, defining
\begin{equation}
    \R(r,b)=\frac{D(r)}{A(r)}-b^2.
    \label{eq:mR}
\end{equation}
Here, $b$ represents the impact parameter of incoming light rays, and $R$ denotes the closest approach to the black hole, determined from the condition $\mathcal{R}(R,b)=0$, which allows expressing $R$ in terms of $b$. As $R\to r_p$, we observe $\Delta\phi\to\infty$, indicating that light rays undergo an infinite number of revolutions around the black hole. This limit defines the critical impact parameter $b_c$, given by the condition $\mathcal{R}(r_p,b_c)=0$, yielding
\begin{equation}
    b_c = \sqrt{\frac{D(r_p)}{A(r_p)}}.
    \label{eq:bc_0}
\end{equation}
Thus, Eq.~\eqref{eq:Deltaphi_0} is valid for $b>b_c$, while light rays with $b=b_c$ remain trapped in spherical orbits, and those with $b<b_c$ are inevitably captured by the black hole. To encompass the entire range of impact parameters, we employ the parametrization~\cite{bozza_strong_2007,tsupko_shape_2022,aratore_constraining_2024}
\begin{equation}
    b=b_c\left(1+\epsilon\right),
    \label{eq:bepsilon}
\end{equation}
where $-1\leq\epsilon<\infty$. Since strong lensing effects are significant for $b\approx b_c$ (or equivalently, $R\approx r_p$), corresponding to trajectories near the photon sphere, the parameter $\epsilon$ is expected to be small. In other words,
\begin{equation}
    \epsilon=\frac{b-b_c}{b_c}\ll 1.
    \label{eq:bepsilon_1}
\end{equation}
Note that, as $R$ approaches $r_p$, then $\mathcal{R}$ approaches $0$, and hence, the integrals in Eq. \eqref{eq:Deltaphi_0} diverge. In Refs. \cite{Bozza:2002,bozza_strong_2007}, a method was developed to address this divergence using logarithmic approximations. Under these approximations, and assuming the source and observer are located at finite positions, the azimuthal angle shift is given by \cite{aratore_constraining_2024}
\begin{equation}
\Delta\phi = -\bar{a} \ln\left(\frac{\epsilon}{\eta_S\eta_O}\right)+\bar{\xi}+\pi,
    \label{eq:Deltaphi_1}
\end{equation}
where the change of variable $\eta = 1-r_p/r$ has been employed. Here,
\begin{equation}
\bar{\xi}=\bar{a}\ln\left(\frac{2\beta_c}{b_c^2}\right)+k_S+k_O-\pi,
    \label{eq:barxi_0}
\end{equation}
and the involved parameters are given by
\begin{subequations}
    \begin{align}
        & \eta_i=1-\frac{r_p}{r_i},\quad i=\{S,O\},\\
        & \bar{a} = r_p\sqrt{\frac{B(r_p)}{A(r_p) \beta_c}},\\
        & \beta_c = \frac{r_p^2\Bigl[D''(r_p)A(r_p)-A''(r_p)D(r_p)\Bigr]}{2A^2(r_p)},\\
        & k_i=\int_0^{\eta_i}g(\eta)\,\mathrm{d}\eta,\label{eq:ki_0}\\
        & g(\eta)=b_c\sqrt{\frac{B(\eta)}{D(\eta)}}\,\frac{1}{\sqrt{\mathcal{R}(\eta,b_c)}}\frac{r_p}{(1-\eta)^2}-\frac{b_c}{\sqrt{\beta_c}}\sqrt{\frac{B(r_p)}{D(r_p)}}\,\frac{r_p}{|\eta|}.
    \end{align}
\end{subequations}
For the metric functions in the line element \eqref{eq:metric_0}, we obtain
\begin{eqnarray}
    g(\eta) &=& -\sqrt{\frac{1}{f_0-(1-\eta)^2 f(\eta)}}-\Biggl\{
    2 r_p^3+2 \lambda ^2 r_p-4 M \left(\lambda ^2+r_p^2\right)+\mathcal{C} \Biggl[\pi  \lambda ^3-2 \lambda  \left(\lambda ^2+r_p^2\right) \arctan\left(\frac{r_p}{\lambda }\right)+\pi  \lambda  r_p^2+2 \lambda ^2 r_p\Biggr]
    \Biggr\}\nonumber\\
   && \times\Biggl\{
    \sqrt{2} f_0 \eta  \Biggl[r_p \Biggl(-12 M \left(\lambda ^2+r_p^2\right)^3+6 r_p^7+18 \lambda ^2 r_p^5+18 \lambda ^4 r_p^3+6 \lambda ^6 r_p\Biggr)\nonumber\\
   && +\mathcal{C} \Biggl(3 \pi  \lambda ^7+3 \pi  \lambda  r_p^6+2 \lambda ^2 r_p^5+9 \pi  \lambda ^3 r_p^4+16 \lambda ^4 r_p^3+9 \pi  \lambda ^5 r_p^2-6 \lambda  \left(\lambda ^2+r_p^2\right)^3 \arctan\left(\frac{r_p}{\lambda }\right)+6 \lambda ^6 r_p\Biggr)\Biggr]^{1/2}
    \Biggr\}^{-1},
    \label{eq:geta_1}
\end{eqnarray}
where $f(\eta)$ follows from substituting $r=r_p/(1-\eta)$ into the lapse function \eqref{eq:lapse_0}, and $f_0=f(0)$ represents the value at the photon sphere. Since an analytical evaluation of the integral \eqref{eq:ki_0} is infeasible, we rely on numerical methods to determine the deflection angle in the strong lensing regime. However, in the limit $\mathcal{C}\to0$, we obtain
\begin{equation}
g_{\mathrm{sbh}}(\eta) = \frac{\sqrt{3}-\sqrt{3-2 \eta }}{\eta\sqrt{3-2 \eta } },
    \label{eq:geta_sbh}
\end{equation}
which leads to $k_i^{\mathrm{sbh}}=-2\ln\left(3+\sqrt{3+18M/r_i}\right)$. Given that for the Schwarzschild black hole (SBH), $\eta_i=1-3M/r_i$, the deflection angle follows as
\begin{equation}
\Delta\phi_{\mathrm{sbh}}-\pi\equiv\alpha_{\mathrm{sbh}} = -\ln\frac{\epsilon}{\eta_S\eta_O}+\xi_{\mathrm{sbh}},
    \label{eq:Deltaphi_SBH}
\end{equation}
where
\begin{subequations}
    \begin{align}
        & \xi_{\mathrm{sbh}} = -\pi+5\ln(6)+k_S^{\mathrm{sbh}}+k_O^{\mathrm{sbh}},
    \end{align}
\end{subequations}
consistent with Ref. \cite{bozza_strong_2007}. In the asymptotic limit $r_i\to\infty$, this reduces to the well-known expression $\alpha_{\mathrm{sbh}}=-\ln\epsilon+\ln\left(216[2-\sqrt{3}]^2\right)-\pi$, as derived in Refs. \cite{Darwin_gravity_1959,Bozza:2002} for strong lensing around an SBH. From this, it follows that $\xi_{\mathrm{sbh}}=-0.4002$, as reported in Ref. \cite{Bozza:2002}. 

\begin{figure}[htp]
    \centering
    \includegraphics[width=8cm]{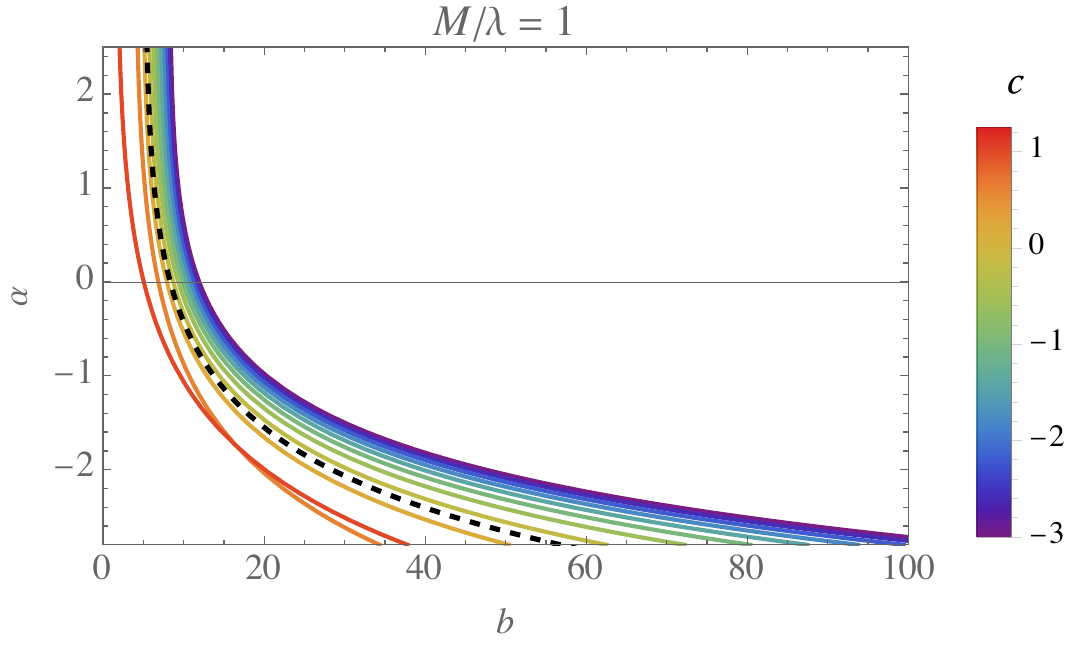} (a)
    \includegraphics[width=8cm]{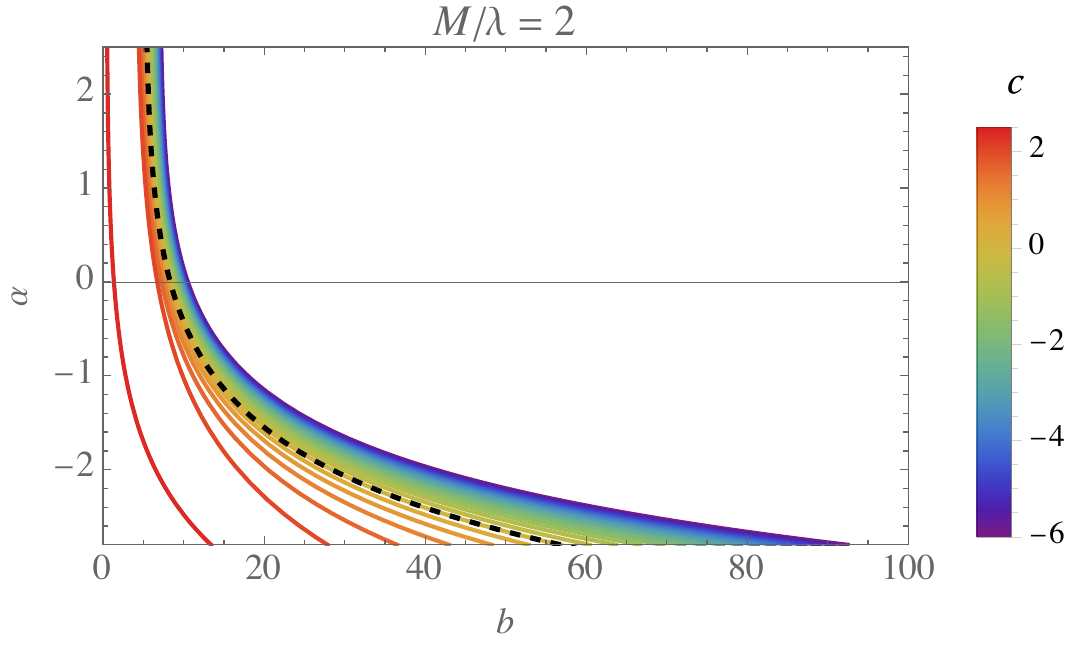} (b)
    \includegraphics[width=8cm]{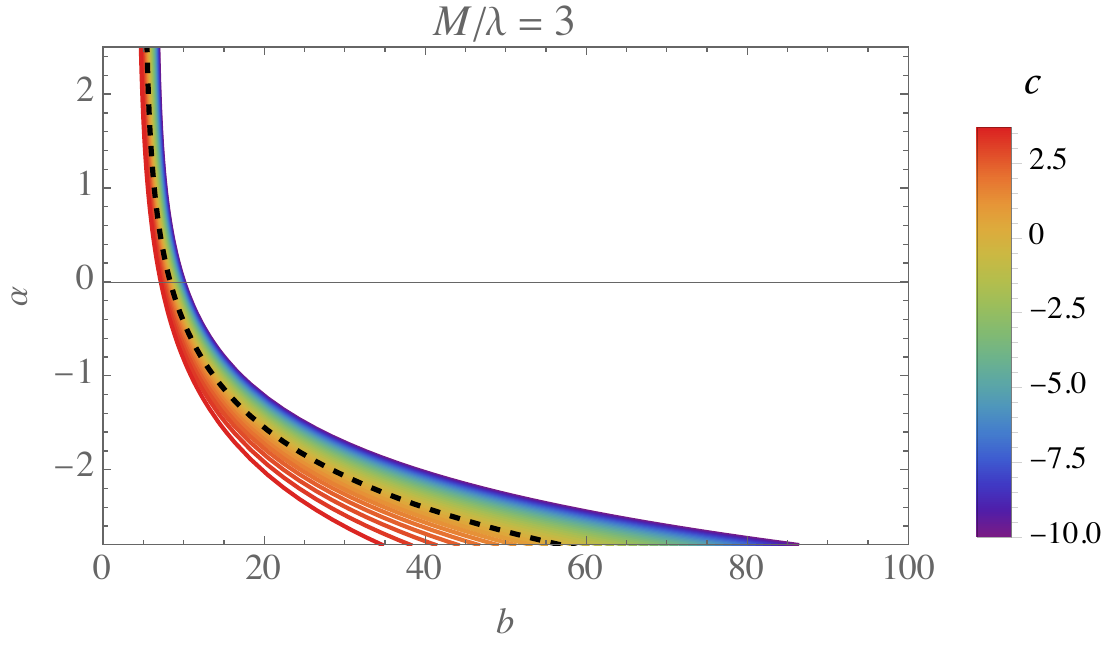} (c)
    \includegraphics[width=8cm]{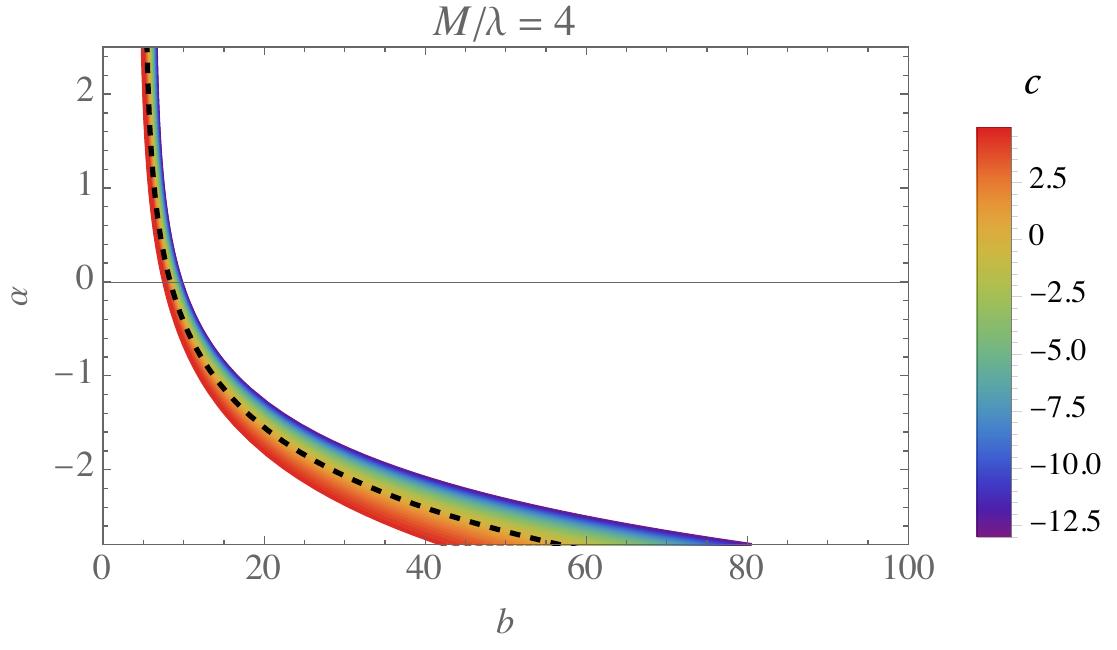} (d)
    \caption{The $b$-profiles of the deflection angle $\alpha$ for the PHBH, plotted for $r_S = r_O = 100$ and different ranges of the $\mc$-parameter, corresponding to the four cases of $M/\lambda = \{1,2,3,4\}$. The range of the $\mc$-parameter in each case is chosen such that $r_p$ possesses real positive values. The dashed curve represents the deflection angle of the SBH. The unit of length is taken as the black hole mass $M$.}
    \label{fig:defangle}
\end{figure}

Figure \ref{fig:defangle} illustrates the deflection angle $\alpha\equiv\Delta\phi-\pi$ for the PHBH, demonstrating its sensitivity to the scalar hair. 
As expected, the diagrams show that the deflection angle diverges at $b = b_c$ for each branch and then decreases rapidly with increasing $b$. Moreover, the rate of decrease in $\alpha$ becomes steeper as the $M/\lambda$ factor increases. Additionally, the deflection angle is larger for $\mc < 0$ than for $\mc > 0$. Consequently, the SBH's deflection angle serves as the upper limit for $\mc > 0$ and the lower limit for $\mc < 0$.

%%%%%%%%%%%%%%%%%
\section{Strong gravitational lensing observables}\label{sec:lensEq}

In this section, we extend our analysis by examining the lens equation and its associated strong lensing observables. Subsequently, we apply these concepts to astrophysical scenarios and derive constraints on the PHBH's $\mc$-parameter based on observational data.

\begin{figure}[h]
    \centering
    \includegraphics[width=8.3cm]{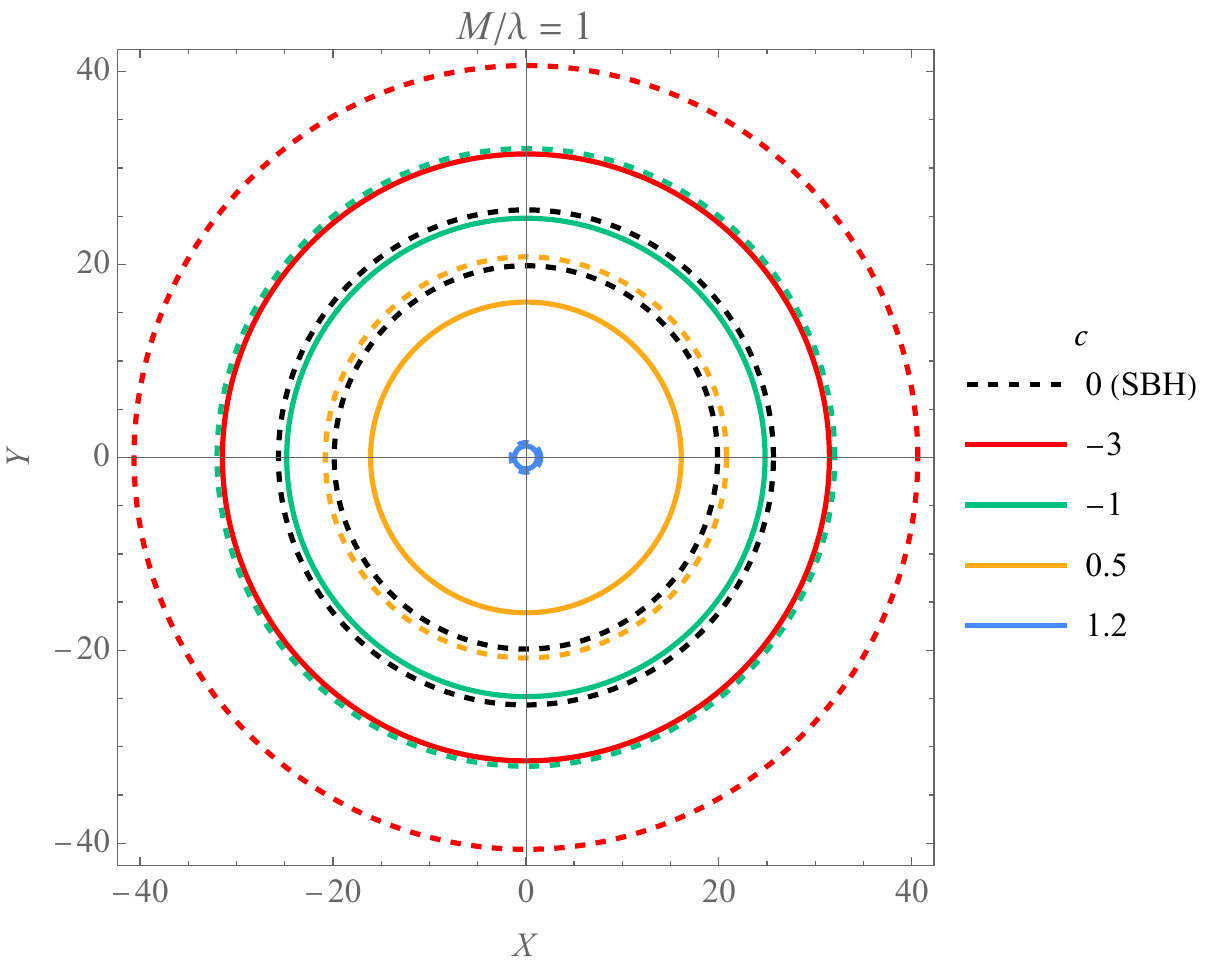} (a)
    \includegraphics[width=8.3cm]{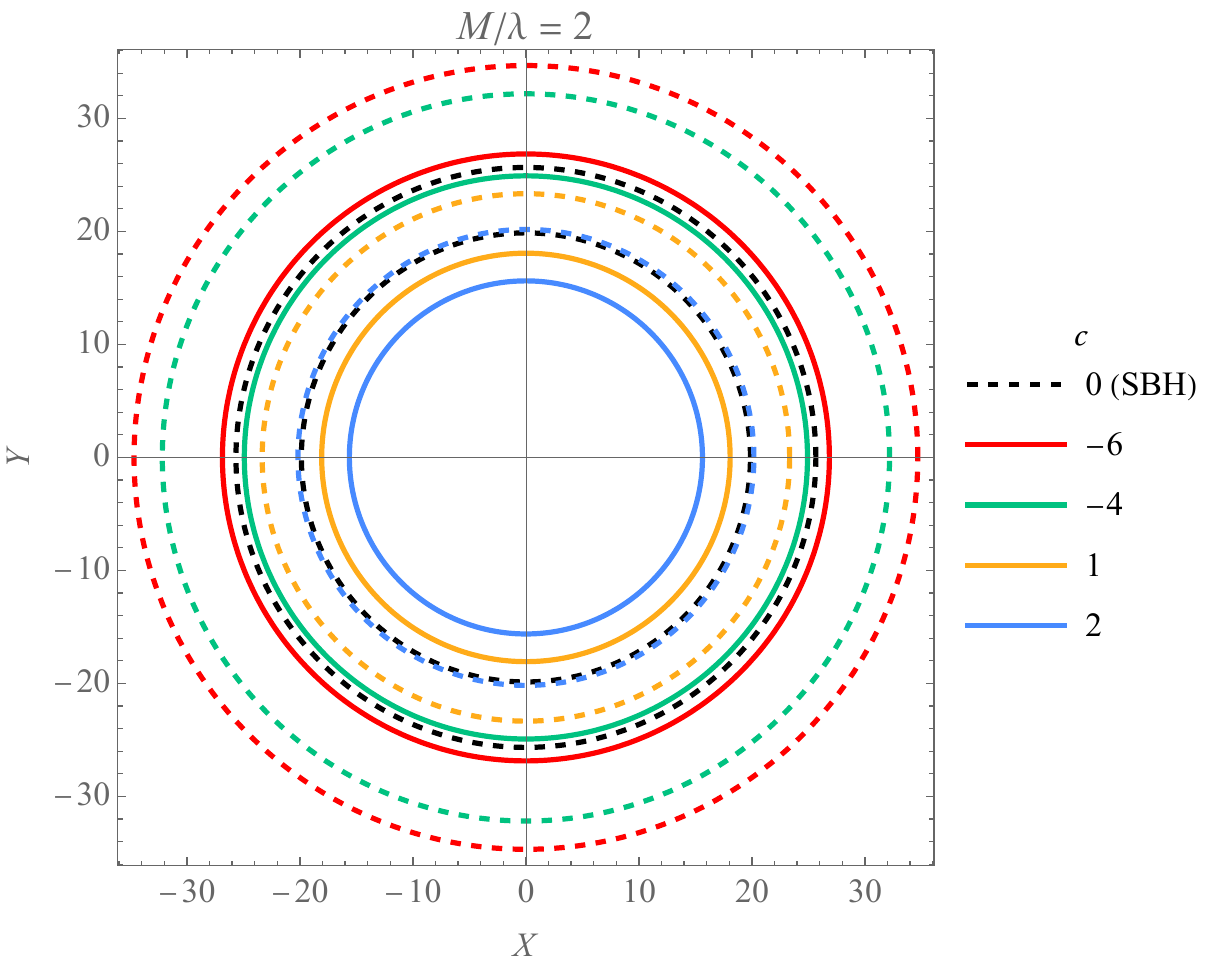} (b)
    \includegraphics[width=8.3cm]{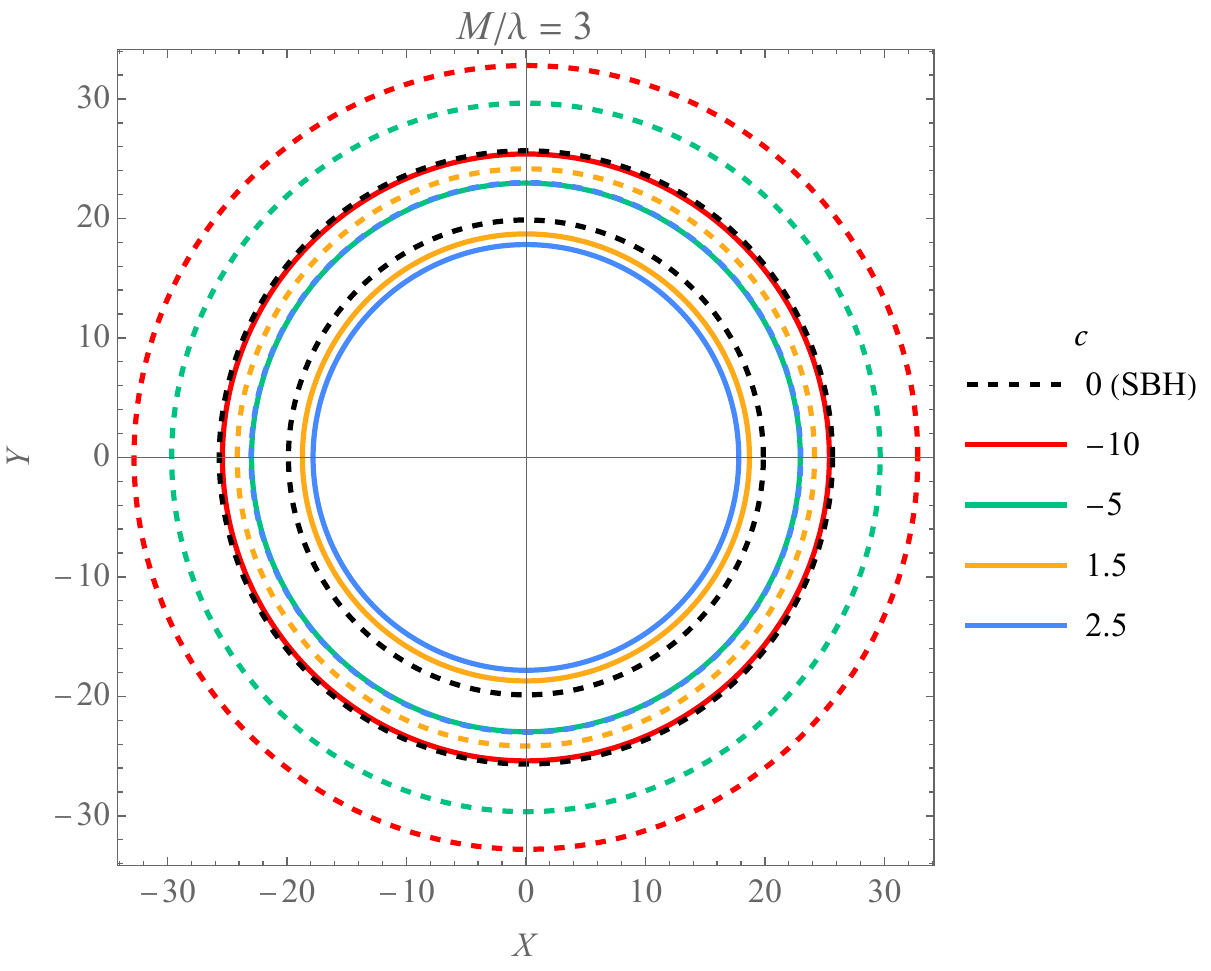} (c)
    \includegraphics[width=8.3cm]{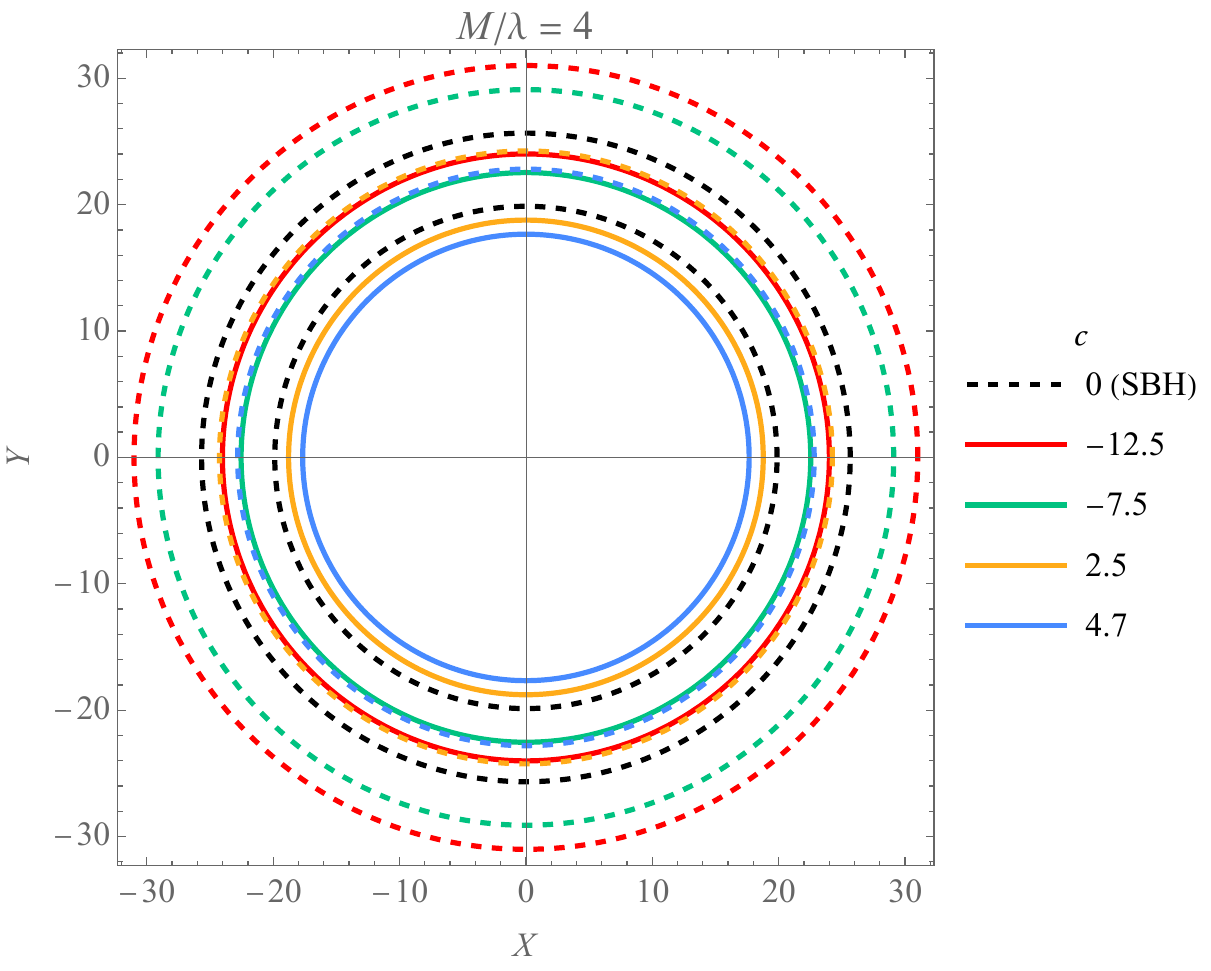} (d)
    \caption{The outermost RERs for M87* and Sgr A* as PHBHs, for the four cases of the $M/\lambda$ ratio, and different values of the $\mc$-parameter corresponding to each of the cases.
    The solid colored lines correspond to M87*, while the dashed colored lines correspond to Sgr A*. 
    The interior and exterior black dashed circles correspond, respectively, to M87* and Sgr A* in the case that they were SBHs. The unit of length has been chosen as the black hole mass $M$, for each of the black holes.}
    \label{fig:Erings}
\end{figure}

To proceed, we express the general lens equation as given in \cite{bozza_strong_2007}
\begin{equation}
\phi_O - \phi_S = \Delta\phi \;\mathrm{mod}\; 2\pi,
    \label{eq:lensEq_0}
\end{equation}
where we set $\phi_O = \pi$ and $\phi_S = 0$. Consequently, from Eq. \eqref{eq:Deltaphi_1}, the positions of the images are given by
\begin{equation}
\epsilon_n = \eta_O \eta_S \exp\left[\frac{\bar{\xi} - 2n\pi}{\ba}\right],
    \label{eq:epsilon_n}
\end{equation}
where $n$, as introduced earlier, represents the number of loops the light rays complete around the black hole. Although exact strong deflection occurs in the limit $n \to \infty$, a good approximation can still be achieved for $n = 1$. For an observer located in an asymptotic region, the angular separation between the image and the black hole is given by $\theta = b/r_O$. From Eq. \eqref{eq:bepsilon}, we obtain $\theta = \theta_c (1 + \epsilon)$, where $\theta_c = b_c / r_O$ denotes the angular radius of the black hole's shadow. In the regime where $r_i \gg r_p$ (considering terms only up to first order in $r_p/r_i$), the total azimuthal shift is given by \cite{Bozza:2002}
\begin{equation}
\Delta\phi = -\ba \ln\left(\frac{r_O \theta}{b_c} - 1\right) + \bar{\xi}.
    \label{eq:Deltaphi_2}
\end{equation}
Within this framework, the lens equation takes the form \cite{Bozza:2001}
\begin{equation}
\psi = \theta - \frac{r_S}{r_{OS}} \Delta\alpha_n,
    \label{eq:lensEq_1}
\end{equation}
where $r_{OS} = r_O + r_S$ is the distance between the source and the observer, $\Delta\alpha_n = \alpha(\theta) - 2n\pi$ represents the offset of the deflection angle after subtracting all loops completed by the photons, and $\psi$ denotes the angular position of the source relative to the black hole. Solving Eq. \eqref{eq:Deltaphi_2} for $\alpha(\theta_n^0) = 2n\pi$ yields
\begin{equation}
\theta_n^0 = \frac{b_c}{r_O} \left(1 + \epsilon_n\right),
    \label{eq:thetan0}
\end{equation}
where $\epsilon_n$ is given in Eq. \eqref{eq:epsilon_n}. By expanding $\alpha(\theta)$ around $\theta_n^0$ and defining $\Delta\theta_n = \theta - \theta_n^0$, we obtain the offset as
\begin{equation}
\Delta\alpha_n = -\frac{\ba r_O}{b_c \epsilon_n} \Delta\theta_n.
    \label{eq:Deltalpha_n}
\end{equation}
Thus, the lens equation can be rewritten as
\begin{equation}
\psi = \theta + \left(\frac{\ba r_S r_O}{b_c \epsilon_n r_{OS}}\right) \Delta\theta_n.
    \label{eq:lensEq_2}
\end{equation}
Considering the condition $r_O \gg b_c$, the angular position of the $n$th relativistic image is given by \cite{Bozza:2002}
\begin{equation}
\theta_n = \theta_n^0 + \frac{b_c \epsilon_n \left(\psi - \theta_n^0\right) r_{OS}}{\ba r_S r_O}.
    \label{eq:theta_n}
\end{equation}
From this equation, it is evident that when $\psi = \theta_n^0$, the image aligns with the source. The sign of $\psi$ determines whether the image appears on the same side ($\psi > 0$) or the opposite side ($\psi < 0$) of the lens. In the scenario where the black hole is nearly aligned with the source and observer ($\psi \approx 0$), and when the observer and lens are positioned equidistantly from the light source ($r_{OS} = r_S = 2 r_O$), light deflection occurs in all directions, leading to the formation of {relativistic Einstein rings (RERs)} \cite{einstein_lens-like_1936, mellier_probing_1999, bartelmann_weak_2001,petters_relativistic_2003,schmidt_weak_2008, guzik_tests_2010}. In this case, Eq. \eqref{eq:theta_n} simplifies to \cite{bozza_time_2004}
\begin{equation}
\theta_n^{E} = \left(1 - \frac{b_c \epsilon_n r_{OS}}{\ba r_S r_O}\right) \theta_n^0.
    \label{eq:theta_nE_0}
\end{equation}
Again, for the case where $r_O \gg b_c$, the angular radius of the $n$th relativistic Einstein ring is given by
\begin{equation}
\theta_n^{E} = \frac{b_c \left( 1 + \epsilon_n \right)}{r_O}.
    \label{eq:thetanE_1}
\end{equation}
To apply the above relations to astrophysical contexts, we now consider the observational data for the supermassive black holes M87* and Sgr A*. Specifically, M87*, with a mass of $(6.5 \pm 0.7) \times 10^9 \, M_\odot$, is located at a distance of $r_O = 16.8 \, \mathrm{Mpc}$ from Earth \cite{the_event_horizon_telescope_collaboration_first_2019, the_event_horizon_telescope_collaboration_first_2019-1}, while Sgr A* has a mass of $4^{+1.1}_{-0.6} \times 10^6 \, M_\odot$ and is situated at a distance of $7.97 \, \mathrm{kpc}$ from Earth \cite{Akiyama:2022, event_horizon_telescope_collaboration_first_2022-1}. Using these parameters in Eq.~\eqref{eq:thetanE_1}, Fig.~\ref{fig:Erings} displays the outermost RERs (corresponding to $n = 1$) for both M87* and Sgr A*, assuming that they are PHBHs. These rings are presented in the celestial coordinate system of an observer on Earth, with coordinates $X$ and $Y$, and are computed for different values of the $M/\lambda$ ratio and for specific ranges of the $\mc$ parameter.
As we can observe, for each fixed positive $\mc$-parameter, an increase in the $M/\lambda$ ratio results in an increase in the radius of the rings. Conversely, for $\mc < 0$, this increase results in a decrease in the rings' radius. Furthermore, for each fixed $M/\lambda$ ratio, an increase in the $\mc$-parameter from its negative to positive values results in a decrease in the rings' radius.

Another important quantity in strong lensing is the magnification of the $n$th relativistic image, which is defined as \cite{Virbhadra:2000,Bozza:2002}
\begin{equation}
\mu_n = \left.\left(\frac{\psi}{\theta} \frac{\ed \psi}{\ed \theta}\right)^{-1}\right|_{\theta_n^0} = \frac{b_c^2 \epsilon_n \left( 1 + \epsilon_n \right) r_{OS}}{\ba \psi r_S r_O^2}.
    \label{eq:mun}
\end{equation}
As we can see, the magnification is inversely proportional to $r_O^2$. Hence, the brightness of the relativistic images is relatively faint. The outermost image is the brightest, and as the order of the images increases, their brightness falls exponentially. However, in the limit as $\psi$ approaches zero (indicating the nearly perfect alignment of the source and the lens), the images can be remarkably magnified. It is important to note that while the outermost images (corresponding to $\theta_1$) remain distinct, higher-order images cluster together around $\theta_\infty \equiv \theta_n |_{n \to \infty}$ \cite{Bozza:2002}, which can be identified as $\theta_\infty = \theta_c$, as mentioned earlier in this section.

The astrophysical applications of strong lensing also depend on two other key observables: the angular separation of the outermost and innermost relativistic images, calculated as
\begin{equation}
s = \theta_1 - \theta_\infty \approx \theta_\infty \epsilon_1,
    \label{eq:s_0}
\end{equation}
and the relative magnification between the outermost relativistic image and the group of inner relativistic images, given by \cite{Bozza:2002}
\begin{equation}
r_{\mathrm{mag}} = \frac{\mu_1}{\sum_{n=2}^\infty \mu_n} = 2.5 \log_{10} \left( \exp \left[ \frac{2\pi}{\ba} \right] \right),
    \label{eq:rmag_0}
\end{equation}
which does not depend on the observer's distance $r_O$.

Now, with these relations at hand, one can estimate the three observables $\theta_\infty$, $s$, and $r_{\mathrm{mag}}$ based on astronomical observations, which demonstrates the impacts of scalar hair on these observables for the PHBH (see Table \ref{tab:1}).
\begin{table*}[htb!]
\resizebox{15cm}{!}{
 \begin{centering}	
	\begin{tabular}{p{2cm} p{2cm} p{2cm} p{2cm} p{2cm} p{2cm} p{2cm}}
\hline\hline
\multicolumn{2}{c}{}&
\multicolumn{2}{c}{M87*} &
\multicolumn{2}{c}{Sgr A*}\\
{$M/\lambda$ } & {$\mc$}& {$\theta_{\infty}$($\mu$as)} & {$s$ ($\mu$as)} & {$\theta_{\infty}$($\mu$as)}  & {$s$ ($\mu$as) } & {$r_{\mathrm{mag}}$} \\ \hline
\hline
\multirow{7}{*}{1}&0.0 (SBH)& 19.847 & 0.025 & 25.629 &0.0321& 6.822\\
& $-3$ & 31.451 & 0.010 & 40.614 & 0.013 & 8.059\\
& $-1$ & 24.781& 0.015 & 32.001 & 0.019 & 7.512\\
& 0.5 & 16.055 & 0.049 & 20.733 & 0.063 & 5.988\\
& 1.2 & 1.183 & 0.002 & 1.528 & 0.002 & 6.809\\
\hline
\multirow{6}{*}{2}
& $-6$ &  26.845 & 0.012 & 34.666 & 0.015 & 7.789 \\
& $-4$ & 24.906 & 0.014 & 32.162 & 0.018 & 7.597\\ 
& 1 & 18.038 & 0.036 & 23.293 & 0.046 & 6.364\\ 
& 2 & 15.546 & 0.078 & 20.075 & 0.101 & 5.349\\ 
\hline
\multirow{6}{*}{3}
& $-10$ & 25.384 & 0.013 & 32.779 &  0.017 & 7.662\\ 
& $-5$ & 22.943 & 0.016 & 29.627 & 0.021 & 7.367 \\ 
& 1.5 & 18.674 & 0.031 & 24.114 &  0.040 & 6.530\\ 
& 2.5 & 17.775 & 0.039 & 22.954 & 0.050 & 6.255 \\ 
\hline
\multirow{6}{*}{4}
& $-12.5$ & 23.998 & 0.015 & 30.989 &  0.019 & 7.510\\ 
& $-7.5$ & 22.520 & 0.017 & 29.081 & 0.022 & 7.310 \\ 
& 2.5 & 18.748 & 0.031 & 24.209 &  0.040 & 6.550\\ 
& 4.7 & 17.623 & 0.041 & 22.760 & 0.053 & 6.190 \\

		\hline\hline
	\end{tabular}
\end{centering}
}	
	\caption{Estimates for the lensing observables, considering the supermassive black holes M87* and Sgr A* as PHBHs, based on the parameter values assumed in Fig. \ref{fig:Erings}.
    The unit of length is chosen as the mass $M$ for each of the black holes.  
    }
    \label{tab:1}
\end{table*} 
We observe that, for any fixed $\mc$-parameter, an increase in the $M/\lambda$ ratio results in a rise in both the brightness ratio $r_{\mathrm{mag}}$ and the image position $\theta_\infty$. On the other hand, when the $M/\lambda$ ratio is fixed, an increase in the $\mc$-parameter from negative to positive values leads to a decrease in both $r_{\mathrm{mag}}$ and $\theta_\infty$. The only exception occurs for the case of $M/\lambda = 1$ and $\mc = 1.2$. As observed in Fig. \ref{fig:Erings}(a) and Table \ref{tab:1}, this case produces smaller RERs compared to other scenarios, and hence, it is considered unreliable within the theory, particularly when observational constraints are applied (see below).

Before concluding this section, it is appropriate to derive some constraints on the scalar hair parameter of black holes using recent observations of M87* and Sgr A*. 

In 2019, the EHT Collaboration released the first horizon-scale image of the supermassive black hole M87*, providing compelling observational evidence for the existence of black holes. Their analysis revealed that the compact emission region had an angular diameter of $\theta_d = (42 \pm 3) \, \mu\text{as}$, along with a central flux depression exceeding a factor of $\gtrsim 10$, which corresponds to the black hole's shadow \cite{eht1, the_event_horizon_telescope_collaboration_first_2019, the_event_horizon_telescope_collaboration_first_2019-1}. In 2022, the EHT Collaboration captured an image of Sgr A*, the supermassive black hole at the center of the Milky Way. This image displayed a distinctive ring with an angular diameter of $\theta_d = (48.7 \pm 7) \, \mu\text{as}$, and showed a deviation from the Schwarzschild shadow characterized by $\delta = -0.08^{+0.09}_{-0.09}$ (VLTI) and $\delta = -0.04^{+0.09}_{-0.10}$ (Keck). Additionally, the EHT results for Sgr A* provided an estimation of the angular diameter of the emission ring, given by $\theta_d = (51.8 \pm 2.3) \, \mu\text{as}$ \cite{Akiyama:2022, event_horizon_telescope_collaboration_first_2022}.  

\begin{figure}[htp]
    \centering
    \includegraphics[width=6.5cm]{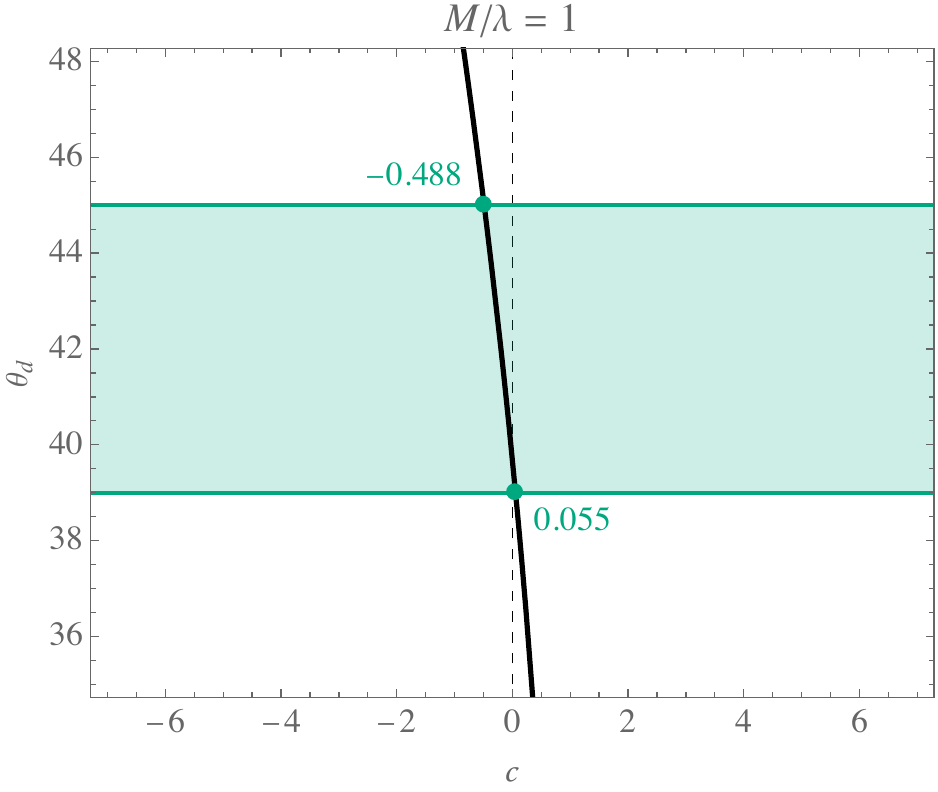} (a)\qquad
    \includegraphics[width=6.5cm]{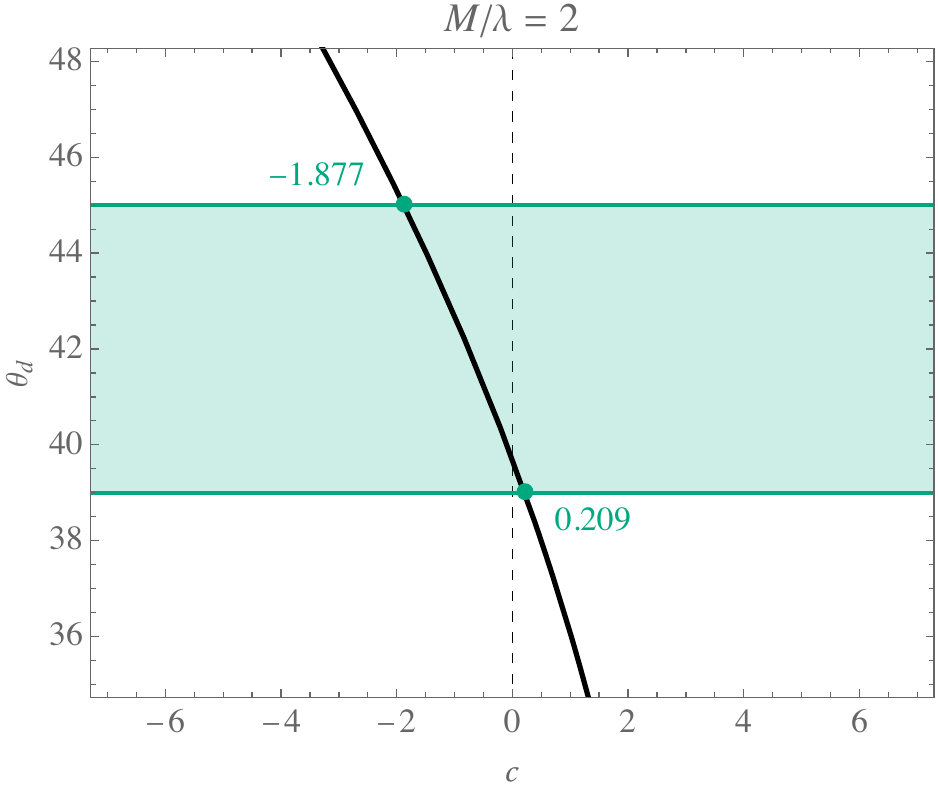} (b)
    \includegraphics[width=6.5cm]{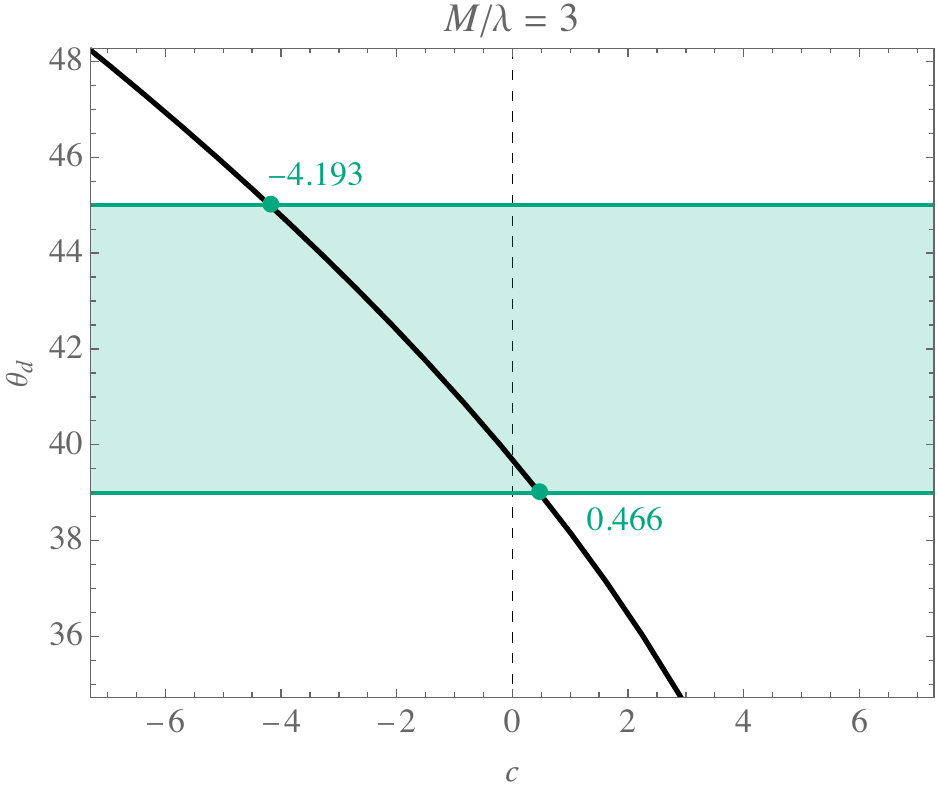} (c)\qquad
    \includegraphics[width=6.5cm]{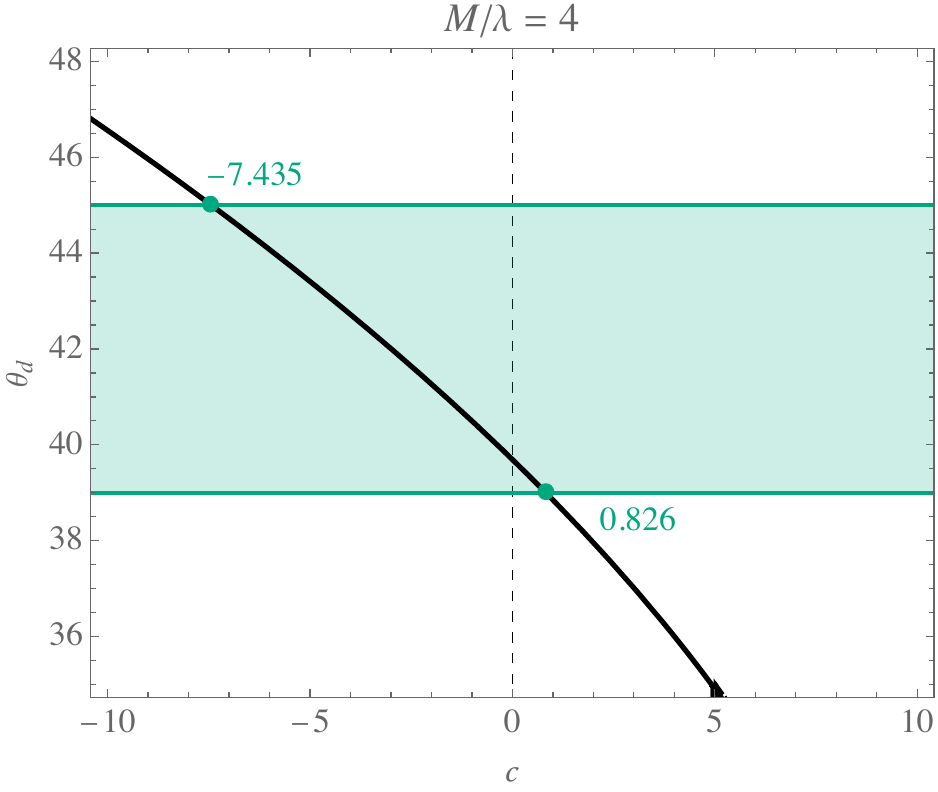} (d)
    \caption{The $\mc$-profile of the theoretical shadow angular diameter ($2\theta_\infty$) of PHBHs (in $\mu$as), compared with the observed shadow diameter $\theta_d$ of M87*, within the $1\sigma$ confidence level, for four different values of the $M/\lambda$ ratio.}
    \label{fig:EHTconstraintsM87_0}
\end{figure}

Using these observational data from the EHT, we model M87* and Sgr A* as PHBHs to constrain the scalar-hair parameter $\mc$ associated with the PHBH. By considering the apparent radius of the photon sphere, $\theta_{\infty}$, as the angular size of the black hole shadow, we derive constraints on $\mc$ at the $1\sigma$ confidence level.  To proceed, we relate $\theta_d = 2 \theta_\infty$, establishing a connection between the observed angular diameter of the shadows and the theoretical shadow angular diameter, as discussed earlier. In Fig. \ref{fig:EHTconstraintsM87_0}, we plot the $\mc$-profile of $2\theta_\infty$ for the four cases of the $M/\lambda$ ratio, for M87*. 
\begin{figure}[htp]
    \centering
    \includegraphics[width=6.5cm]{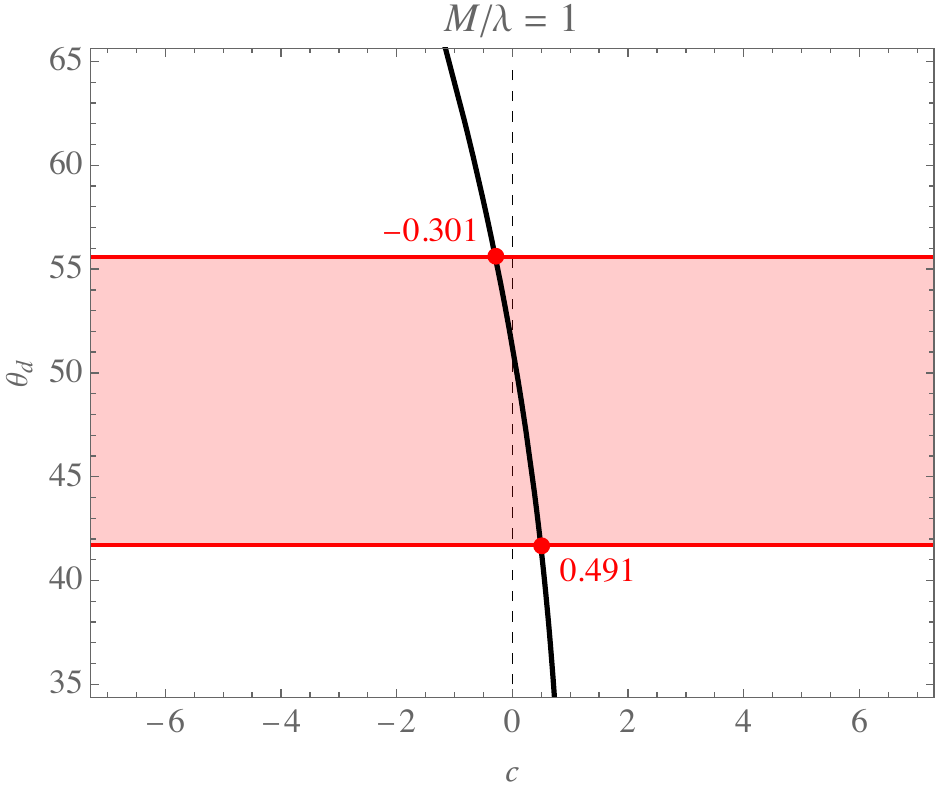} (a)\qquad
    \includegraphics[width=6.5cm]{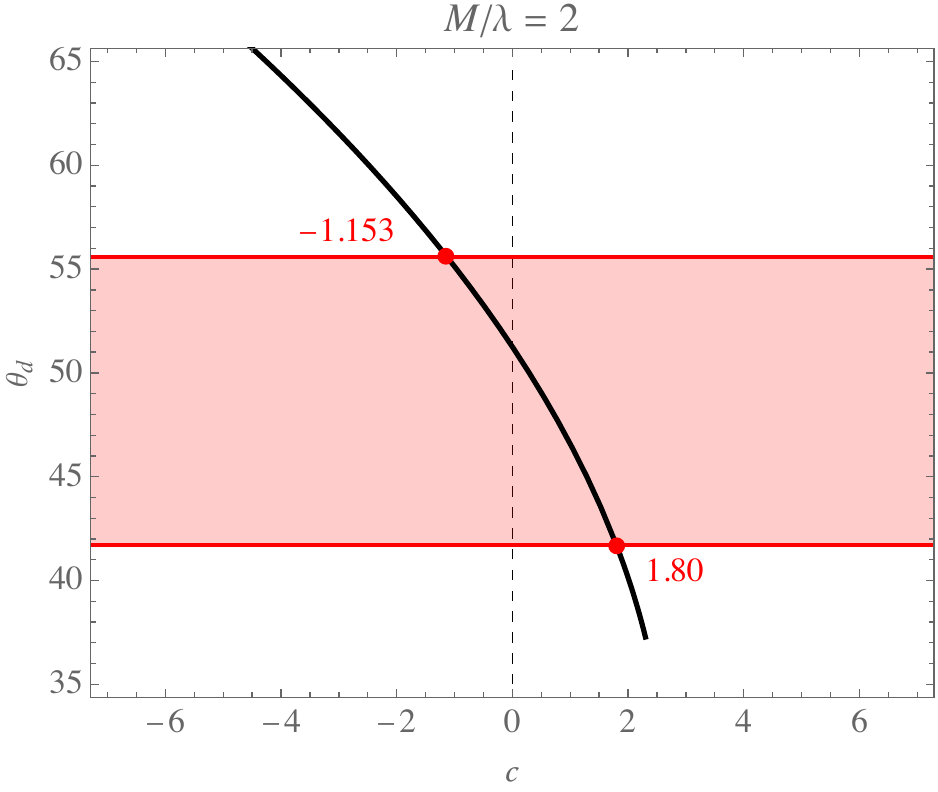} (b)
    \includegraphics[width=6.5cm]{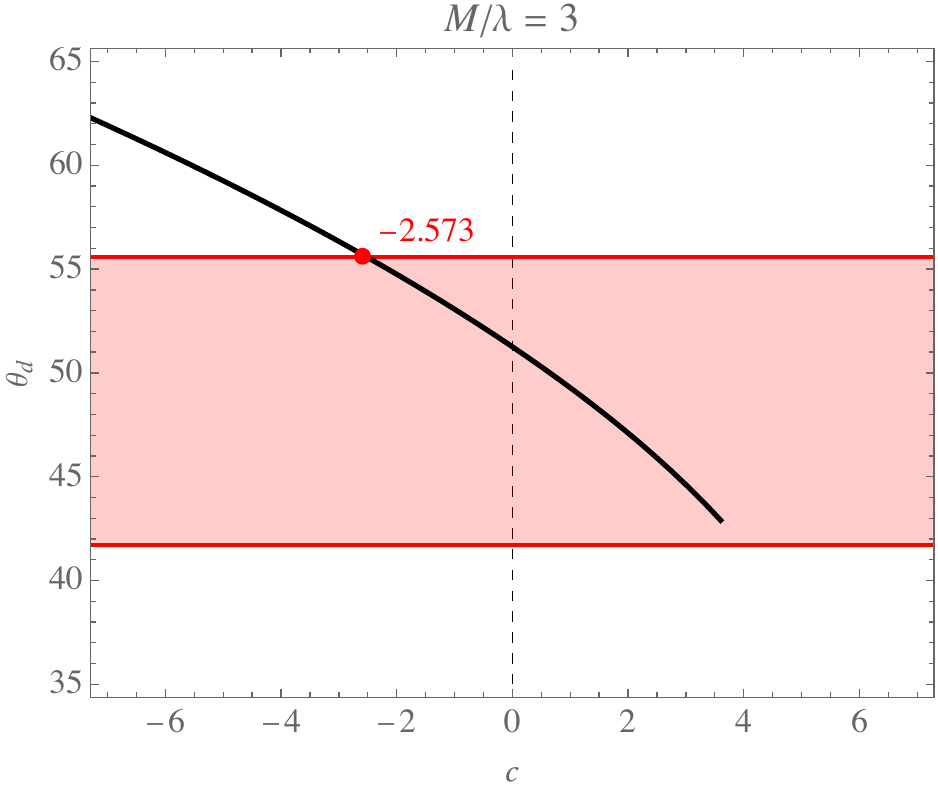} (c)\qquad
    \includegraphics[width=6.5cm]{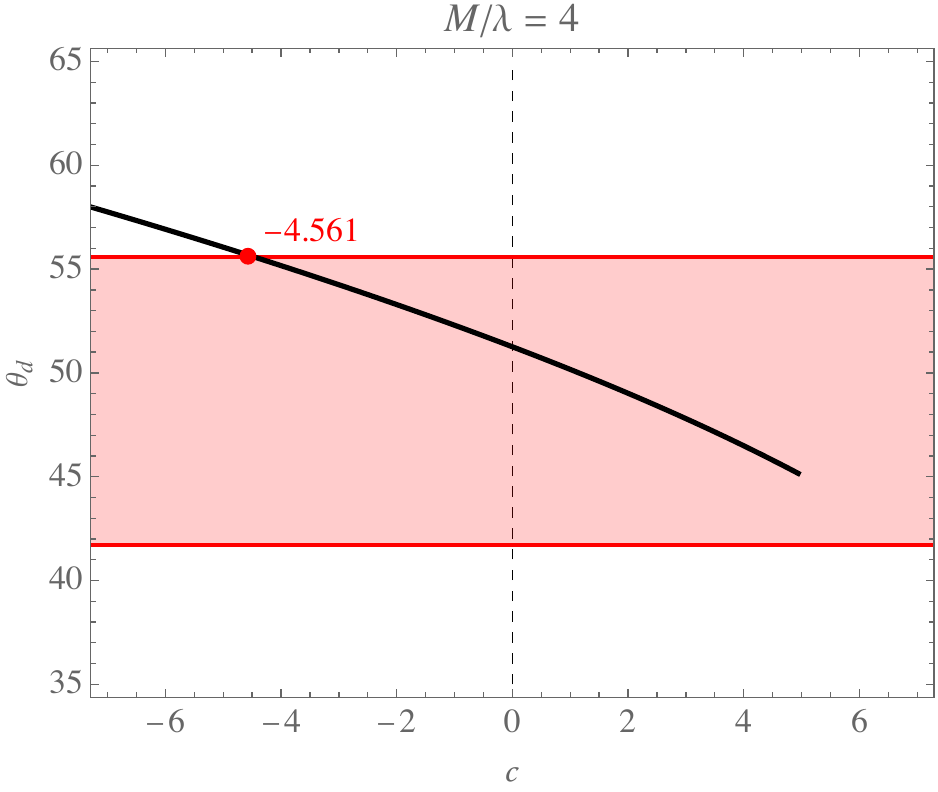} (d)
    \caption{The $\mc$-profile of the theoretical shadow angular diameter ($2\theta_\infty$) of PHBHs (in $\mu$as), compared with the observed shadow diameter $\theta_d$ of Sgr A*, within the $1\sigma$ confidence level, for four different values of the $M/\lambda$ ratio.}
    \label{fig:EHTconstraintsSgrA_0}
\end{figure}
For the case of Sgr A*, we use the averaged shadow angular diameter, which has been obtained by independent algorithms from the EHT observations. In this context, it has been revealed that $\theta_d$ for Sgr A*, resides in the range $46.9\,\mathrm{\mu as}$ -- $50\,\mathrm{\mu as}$, which withing the $1\sigma$ uncertainty, covers the interval $41.7\,\mathrm{\mu as}\leq\theta_d\leq 55.6\,\mathrm{\mu as}$ \cite{event_horizon_telescope_collaboration_first_2022}. In Fig. \ref{fig:EHTconstraintsSgrA_0}, we have considered this confidence range, to compare the theoretical shadow diameter with the observed one, in order to constrain the $\mc$-parameter for four vales of the $M/\lambda$ ratio as in Fig. \ref{fig:EHTconstraintsM87_0}.  

In Table \ref{tab:2}, we summarize the constraints obtained so far on the scalar hair-related parameter, $\mc$, based on strong gravitational lensing by the black hole.
\begin{table}[htp]
	\centering
	\begin{tabular}{cccccccccccc}
		\toprule
		 &\multicolumn{2}{c}{$M/\lambda=1$}& &\multicolumn{2}{c}{$M/\lambda=2$}& &\multicolumn{2}{c}{$M/\lambda=3$}& &\multicolumn{2}{c}{$M/\lambda=4$}\\
		\cmidrule{2-3} \cmidrule{5-6} \cmidrule{8-9} \cmidrule{11-12}
		
		{} & {upper} & {lower} &  {} & {upper} & {lower}  & {} & {upper} & {lower}&  {} & {upper} & {lower}\\
		\midrule
		$\text{M87*}$ & $-0.48762$ & $0.05525$ & & $-1.87733$ & $0.20945$ & & $-4.19312$ & $0.46632$ & & $-7.43517$ & $0.82593$\\
		$\text{Sgr A*}$ & $-0.30069$ & $0.49062$ & & $-1.15299$ & $1.79974$ & & $-2.57314$ & - & & $-4.56132$ & -\\
		\bottomrule
	\end{tabular}
 \caption{The allowable $\mc$-parameter values, obtained from the curves in Figs. \ref{fig:EHTconstraintsM87_0} and \ref{fig:EHTconstraintsSgrA_0}, corresponding to the black hole shadow angular diameter $\theta_d=2\theta_\infty$ that aligns with the EHT observations of M87* and Sgr A* within the $1\sigma$ confidence intervals.}
 \label{tab:2}
\end{table}
These new constraints on the $\mc$-parameter can be compared with those reported in Ref. \cite{erices_thermodynamic_2025}. As seen, for the constraints derived from the EHT data for M87*, the values presented in Table \ref{tab:2} fall within the intervals reported in the aforementioned reference. Therefore, strong lensing imposes more stringent constraints on the black hole scalar hair parameter than those derived from the shadow diameter based on the critical impact parameter, since the former directly depends on the black hole's distance from the observer, i.e., $r_O$.  In contrast, for Sgr A*, the constraints listed in Table \ref{tab:2} differ from those in Ref. \cite{erices_thermodynamic_2025}. This discrepancy arises because, in the present study, we used the averaged shadow diameter within the interval $\theta_d \in (41.7\,\mu\text{as}, 55.6\,\mu\text{as})$, whereas Ref. \cite{erices_thermodynamic_2025} employed an angular diameter of $\theta_d = (48.7 \pm 7)\,\mu\text{as}$.

In the next section, we present an alternative approach to constraining the black hole parameters, based on the diameter of the direct disk emission, as reported by the EHT. This diameter is related to the zeroth-order lensed image, which is theoretically inferred from strong gravitational lensing.

%%%%%%%%%%%%%%%%%
\section{Impact of the primary scalar hair on higher-order accretion disk images}\label{sec:accImages}

In this section, we advance the analysis of the strong lensing by the PHBH by assuming that the black hole is illuminated by an accretion disk. We examine the shape of the disk's images of various orders and explore how the primary hair affects the shape of higher-order images. Additionally, we assess the sensitivity of the black hole model to the primary hair. This analysis is primarily based on the recent studies in Refs. \cite{tsupko_shape_2022,aratore_constraining_2024}.

As discussed in Sect. \ref{sec:deflection}, the order $n$ characterizes the properties of the lensed images of a light source around the black hole, as observed by a distant observer. It is commonly related to the number of half-orbits performed by photons around the black hole \cite{johnson_universal_2020,gralla_lensing_2020,gralla_shape_2020,pesce_toward_2021,wielgus_photon_2021,broderick_measuring_2022,paugnat_photon_2022,vincent_images_2022,ayzenberg_testing_2022}. In this context, the primary image, denoted by $n=0$, is a direct lensed image of the source, assumed to be a thin accretion disk, and is generated by the deflected trajectories that do not closely approach the photon sphere. 
\begin{figure}[t]
    \centering
    \includegraphics[width=8cm]{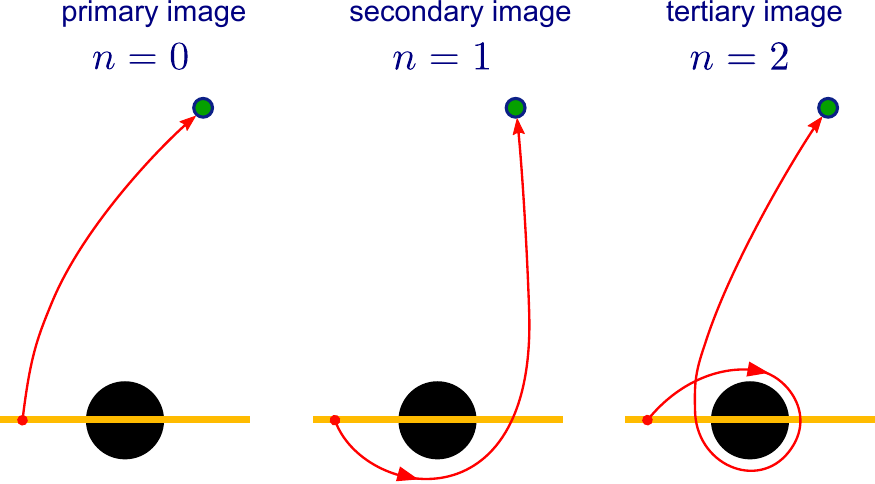}
    \caption{The formation of the first three images of an accretion disk, with images labeled according to the number of half-orbits $n$.
    The figure is taken from Ref. \cite{tsupko_shape_2022}.}
    \label{fig:rings}
\end{figure}
For $n=1$, the secondary image is formed, representing the demagnified image of the far side of the disk. For $n \geq 2$, higher-order photon rings are generated, corresponding to the tertiary image and beyond (see Fig. \ref{fig:rings}). In this section, we investigate the impact of the primary hair component of the PHBH on these three images of the black hole, assuming it is illuminated by a thin accretion disk.

\begin{figure}[htp]
    \centering
    \includegraphics[width=11cm]{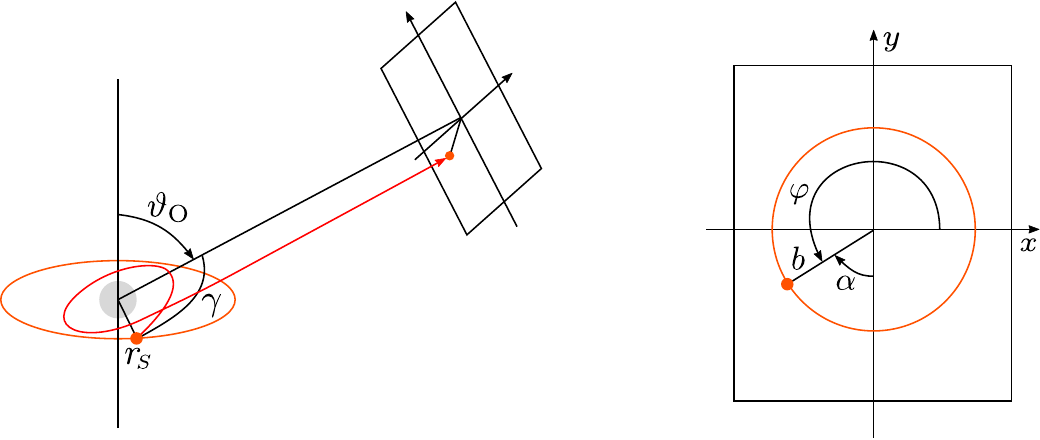}
    \caption{Parametrization of the observer's screen for a radiating ring of radius $r_S$ in the equatorial plane of the black hole. The left panel shows an observer with an inclination angle $\vartheta_O$, with rays responsible for the formation of the tertiary image ($n=2$). The angle $\gamma$ measures the angle between the source and the observer's line of sight within the light ray's plane. The right panel shows the observer's screen parametrized by the polar coordinates $b$ and $\alpha$. This latter angle will be transformed to $\varphi$ in this paper.
    The figure is taken from Ref. \cite{tsupko_shape_2022}.}
    \label{fig:geometry}
\end{figure}

Now, to introduce the mathematical tools used in this analysis, consider the diagram in Fig. \ref{fig:geometry}, which shows a thin emission ring as observed on the screen of an observer located at an arbitrary angle $\vartheta_O$ and at a large radial distance from the black hole's center.
As shown in the figure, the observer's screen is characterized by the impact parameter $b$ and the polar coordinate $\varphi$, which is the transformed form of the angle $\alpha$ presented in Ref. \cite{Luminet:1979nyg}. Based on the mathematical methods outlined in Refs. \cite{tsupko_shape_2022,aratore_constraining_2024}, the shape of the images can be determined from the polar behavior of the function $b(\varphi)$.  To proceed, let us recall that by inverting Eq. \eqref{eq:Deltaphi_1}, we obtain the relation
\begin{equation}
b=b_c\left(1+F(r_S)F(r_O)\exp\left[-\frac{\Delta\phi}{\ba}\right]\right),
    \label{eq:bphi_0}
\end{equation}
where
\begin{equation}
F(r_i)=\eta_i\sqrt{\frac{2\beta_c}{b_c^2}}\,\exp\left[\frac{k_i}{\ba}\right].
    \label{eq:Fi}
\end{equation}
Now, to retrieve the image order $n$, and with $\gamma$ being the convex angle between the arrival and departure directions of the light rays (as shown in Fig. \ref{fig:geometry}), the change in the azimuth angle satisfies the following conditions
\begin{equation}
\Delta\phi=\left\{\begin{array}{cc}
    n\pi+\gamma & \text{for even $n$}, \\
    (n+1)\pi-\gamma & \text{for odd $n$}. 
\end{array}
\right.
    \label{eq:Deltaphi_4}
\end{equation}
As observed, for $n=0$, no half-orbits occur around the black hole, and the light rays do not intersect the equatorial plane (the accretion disk) after their departure. In contrast, for $n \geq 1$, photons perform half-orbits around the black hole, leading to the formation of additional images on the observer's screen. For $n \geq 2$, as photons complete a full loop, higher-order photon rings are generated, which, as shown in Ref. \cite{bozza_strong_2007}, are known as relativistic images.  If the observer is located in the northern hemisphere (i.e., $0 \leq \vartheta_O \leq \pi/2$) and the emission ring of radius $r_S$ is assumed to be thin and positioned on the equatorial plane (i.e., $\vartheta_S = \pi/2$ and $0 \leq \varphi_S \leq 2\pi$), the impact parameter in Eq. \eqref{eq:bphi_0} becomes a function of $\gamma$, which now varies within the range $\pi/2 - \vartheta_O \leq \gamma \leq \pi/2 + \vartheta_O$.  

Since the spacetime is asymptotically flat, we can treat $b$ as the radial coordinate on the observer's screen. Moreover, the angle $\gamma$ can be related to the polar angle $\varphi$ on the observer's screen using Luminet's formula \cite{Luminet:1979nyg}, which was modified by Tsupko in Ref. \cite{tsupko_shape_2022} as
\begin{equation}
\gamma=\left\{\begin{array}{cc}
    \pi-\arccos\left(\frac{\sin\varphi}{\sqrt{\sin^2\varphi+\cot^2\vartheta_O}}\right) & \text{for even $n$}, \\
    \arccos\left(\frac{\sin\varphi}{\sqrt{\sin^2\varphi+\cot^2\vartheta_O}}\right) & \text{for odd $n$},
\end{array}
\right.
    \label{eq:gamma_varphi_0}
\end{equation}
where $0\leq\arccos\leq\pi$, and $0\leq\varphi\leq 2\pi$. Note that, from Eq. \eqref{eq:gamma_varphi_0}, we can derive a single expression for the shift in the azimuth angle as \cite{tsupko_shape_2022}
\begin{equation}
\Delta\phi=(n+1)\pi-\arccos\left(\frac{\sin\varphi}{\sqrt{\sin^2\varphi+\cot^2\vartheta_O}}\right).
    \label{eq:Deltaphi_5}
\end{equation}
Thus, the impact parameter in Eq. \eqref{eq:bphi_0} can be expressed as
\begin{equation}
b_n(\varphi)=b_c\left\{
1+\frac{2\eta_O\eta_S\beta_c}{b_c^2}\,\exp\left[\frac{k_O+k_S-(n+1)\pi}{\ba}\right]\times\exp\left[\frac{1}{\ba}\arccos\left(\frac{\sin\varphi}{\sqrt{\sin^2\varphi+\cot^2\vartheta_O}}\right)\right]
\right\}.
    \label{eq:bphi_1}
\end{equation}
In the limit where $r_O\rightarrow\infty$ (or equivalently $\eta_O\rightarrow1$), the above equation reduces to the form introduced in Ref. \cite{aratore_constraining_2024}. The expression in Eq. \eqref{eq:bphi_1} determines the shape of the $n$th-order image.

Before proceeding with the derivation of the shapes of the images of various orders, it is important to note that the formation of the accretion disk, as the light source, is strictly based on particles traveling along the innermost stable circular orbits (ISCO). In fact, the ISCO defines the smallest stable circular orbit, beyond which matter spirals into the black hole. Observations and simulations confirm that the inner edge of the accretion disk coincides with this boundary, where angular momentum transport ceases. 

To determine the ISCO for the PHBH, we consider the Lagrangian dynamics governing the motion of massive particles in the spacetime by defining the Lagrangian
\begin{equation}
\mL(\bm{x}, \dot{\bm{x}}) = \frac{1}{2} g_{\mu\nu} \dot{x}^\mu \dot{x}^\nu,
    \label{eq:Lagrangian}
\end{equation}
where $\dot{\bm{x}}\equiv\ed \bm{x}/\ed\tau$, with $\tau$ being the affine parameter of the geodesic curves. For the general spacetime \eqref{eq:metr0}, this leads to the expression
\begin{equation}
\mL(\bm{x},\dot{\bm{x}})=\frac{1}{2}\left[-f(r)\dot t^2+\frac{\dot r^2}{f(r)}+r^2\left(\dot\theta^2+\sin^2\theta\,\dot\phi^2\right)\right].
    \label{eq:Lagrangian_1}
\end{equation}
Without loss of generality, we restrict ourselves to the equatorial plane by setting $\theta=\pi/2$. The constants of motion are then defined as
\begin{eqnarray}
    && E=f(r)\dot t,\label{eq:E}\\
    && L=r^2\dot\phi,\label{eq:L}
\end{eqnarray}
which correspond to the energy and angular momentum of the test particles, respectively. Accordingly, the orbit of massive particles is characterized by the Hamilton-Jacobi equation $2\mathcal{L}=-1$. The corresponding effective potential is given by
\begin{equation}
\mathcal{V}(r)=-r^4\left(\frac{E_t^2}{L_t^2}-\frac{f(r)}{L_t^2}-\frac{f(r)}{r^2}\right),
    \label{eq:Veff}
\end{equation}
where $E_t$ and $L_t$ are the energy and angular momentum of the massive test particles. The necessary condition for the formation of accretion disks is the presence of at least one minimum in the effective potential's radial profile, a condition that supports the formation of planetary bound orbits. Therefore, the ISCO can be determined by the mutual condition $\mathcal{V}(r)=0=\mathcal{V}'(r)$. The radius of the ISCO, denoted by $r_c$, satisfies the relation \cite{Guo:2022}
\begin{equation}
r_c = \frac{3f(r_c) f'(r_c)}{2f'(r_c)^2-f(r_c) f''(r_c)},
    \label{eq:rc}
\end{equation}
in the exterior geometry of the black hole. This radius corresponds to the inner edge of the accretion disk, and as one moves away from the black hole, particles appear to move along Keplerian bound orbits. By incorporating the lapse function in Eq. \eqref{eq:lapse_0} into the above conditions, one can determine $r_c$ for the PHBH, which depends on the model parameters and can serve as one of the important source distances, located at $r_S=r_c$.

\begin{figure}[h]
    \centering
    \includegraphics[width=7cm]{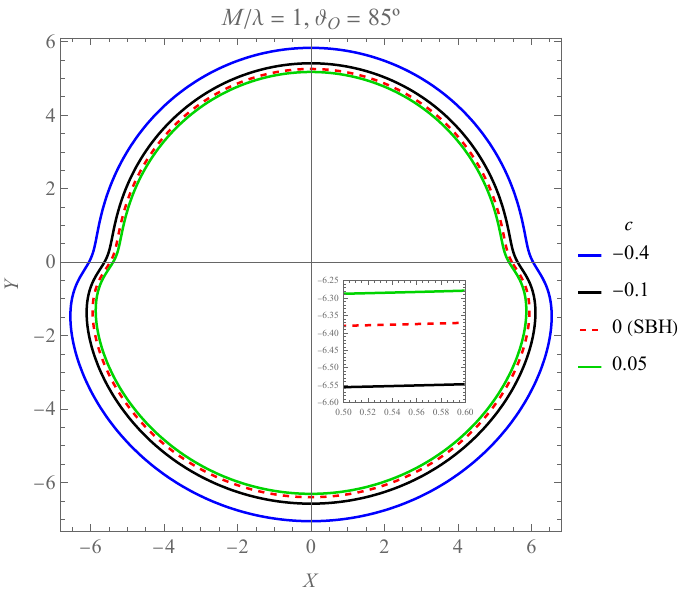} (a)
    \includegraphics[width=7cm]{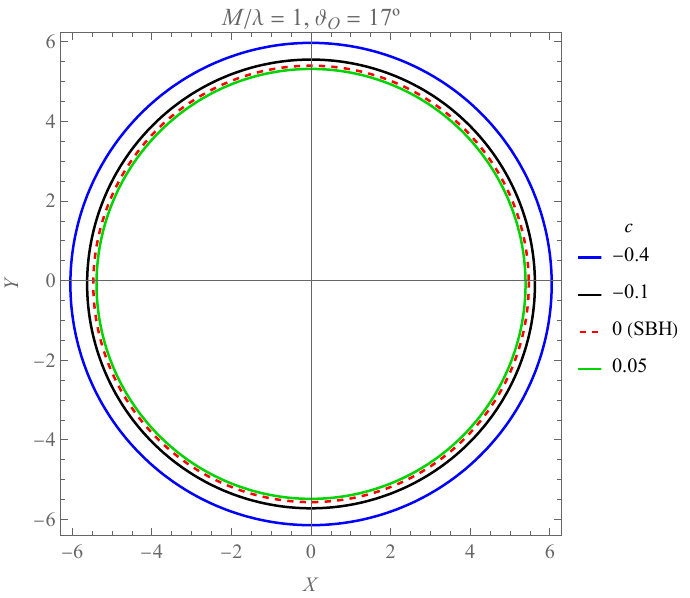} (b)
    \includegraphics[width=7cm]{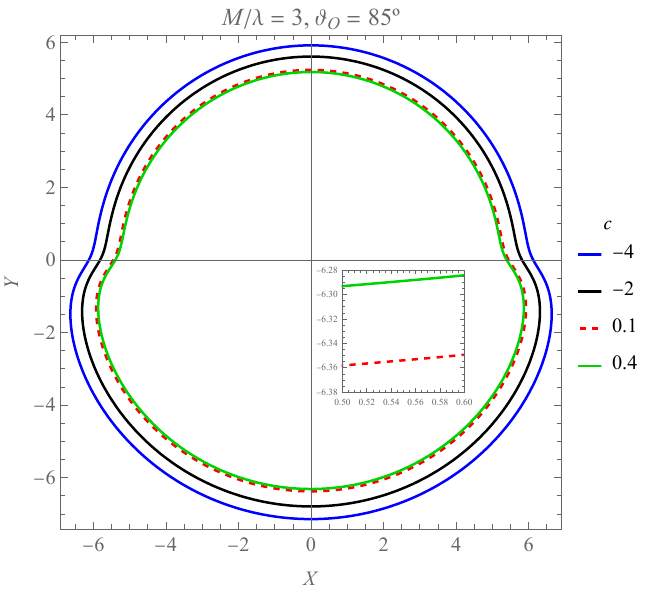} (c)
    \includegraphics[width=7cm]{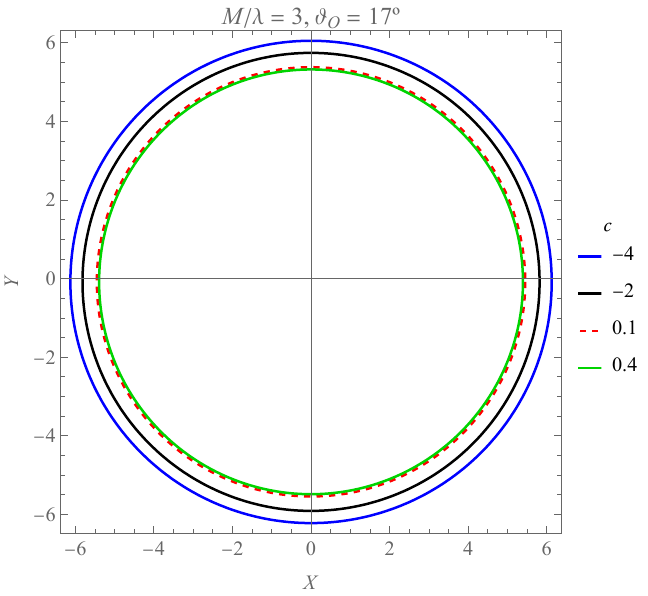} (d)
    \caption{The secondary images of M87* as a PHBH, viewed from two different inclinations $\vartheta_O$ in the observer's screen within the $X$-$Y$ plane. These are plotted for two values of the $M/\lambda$ ratio and various values of the $\mc$-parameter, in accordance with the constraints provided in Table \ref{tab:2}, considering $r_S=r_c$ for each case. 
    The corresponding distances to the source, from the largest rings to the smallest, are $7.021, 6.273, 6.0,$ and $5.858$ for $M/\lambda=1$, and $7.170, 6.618, 5.967$ and $5.866$ for $M/\lambda=3$.
    The unit of length along the axes is the black hole mass $M$.}
    \label{fig:b1_M87images_c}
\end{figure}

\begin{figure}[h]
    \centering
    \includegraphics[width=7cm]{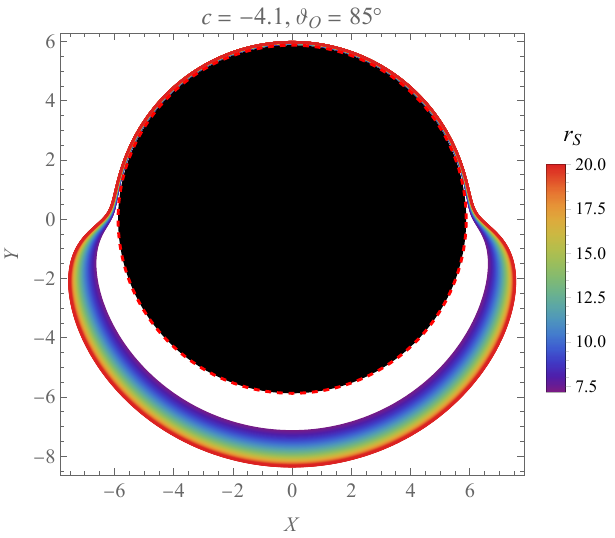} (a)
    \includegraphics[width=7cm]{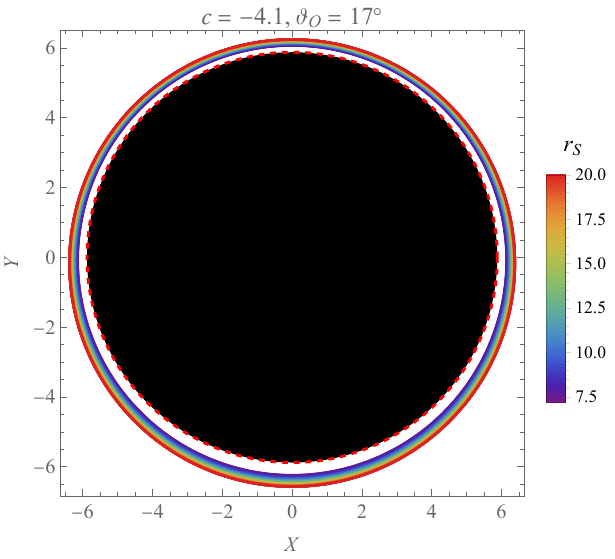} (b)
    \includegraphics[width=7cm]{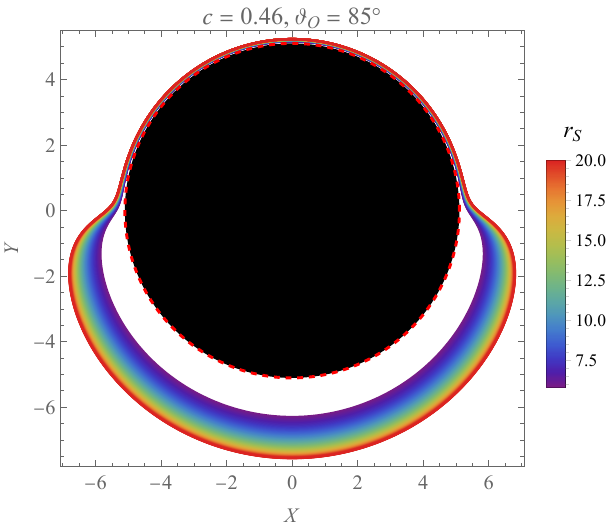} (c)
    \includegraphics[width=7cm]{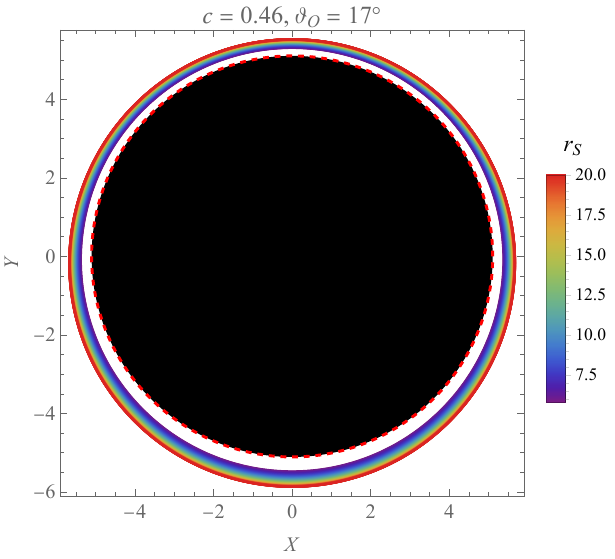} (c)
    \caption{The behavior of the secondary images of M87* as a PHBH with respect to changes in $r_S$, viewed from three different inclinations $\vartheta_O$ in the observer's screen within the $X$-$Y$ plane, plotted for a $M/\lambda=3$ ratio and various values of the $\mc$-parameter, as per the constraints provided in Table \ref{tab:2}. The black disk in the center represents the black hole shadow and has a diameter of $d_{\mathrm{sh}}=2b_\infty=2b_c$, with the values $d_{\mathrm{sh}} = 11.754$ for $\mc=-4.1$, and $d_{\mathrm{sh}} = 10.213$ for $\mc=0.46$.  
    The unit of length along the axes is the black hole mass $M$.}
    \label{fig:b1_M87images_rS}
\end{figure}

Note that the apparent shape of the secondary image is given by
\begin{equation}
b_{\mathrm{sec}}(\varphi)=b_c\left\{
1+\frac{2\eta_O\eta_S\beta_c}{b_c^2}\,\exp\left[-\frac{2\pi-(k_O+k_S)}{\ba}\right]\times\exp\left[\frac{1}{\ba}\arccos\left(\frac{\sin\varphi}{\sqrt{\sin^2\varphi+\cot^2\vartheta_O}}\right)\right]
\right\},
    \label{eq:bphi_1}
\end{equation}
which corresponds to the secondary image of the accretion disk. As the order of the images increases, they approach the boundary of the black hole shadow, located at the radius $b_\infty=b_c$. Therefore, for the secondary image, there remains a gap from the boundary of the shadow. This is demonstrated in Fig. \ref{fig:b1_M87images_c}, where the shape of the secondary images of the accretion disk is shown by setting $r_S=r_c$, in different scenarios for the $\mc$-parameter and $M/\lambda$ ratio, while M87* is considered as a PHBH observed from different inclination angles.
As observed from the diagrams, the rings become increasingly circular as the inclination angle decreases. We have also plotted in Fig. \ref{fig:b1_M87images_rS}, the behavior of the secondary images of M87* with respect to changes in the source distance $r_S$ for different inclination angles, while keeping $M/\lambda$ and $\mc$ fixed.
As observed from the diagrams, the $b_{\mathrm{sec}}$ images form a ring outside the black hole shadow, confined by the $b_\infty=b_c$ image. In images with small inclination angles, this ring is clearly distinct from the inner shadow. In what follows, we exploit this characteristic to derive new constraints on the primary hair-related parameter of the PHBH.

%%%%%%%%%%%%%
\subsection{New constraints on the black hole's primary scalar hair}

The GRMHD simulations, based on observational data from the Event Horizon Telescope (EHT) for M87*, reveal a photon ring with an angular diameter of $\theta_d=(42\pm3)\,\mathrm{\mu as}$, located outside the interior shadow \cite{eht1}. An inclination of $17^\circ$ between the approaching jet and the line of sight was adopted, as stated in Ref. \cite{walker_structure_2018}. This confirms that the aforementioned ring is distinct from the actual black hole shadow, which is bounded by the photon sphere characterized by $b_\infty = b_c$. As observed in the diagrams in Fig. \ref{fig:b1_M87images_rS}, for $\vartheta_O = 17^\circ$, the $b_{\mathrm{sec}}$ image appears nearly as a thick ring. Consequently, one can theoretically associate this ring (formed by the $b_\mathrm{sec}$ images) with the photon ring, which marks the boundary of the shadow of M87*, assuming it to be a PHBH. It is important to note that in the published image of M87*, this ring is dominated by the direct emission from the accretion disk (i.e., the primary image in Fig. \ref{fig:rings}) and is therefore not visible. Thus, the generated ring, which borders the shadow, results from analyzing the data based on the GRMHD simulations.

\begin{figure}[htp]
    \centering
    \includegraphics[width=6.5cm]{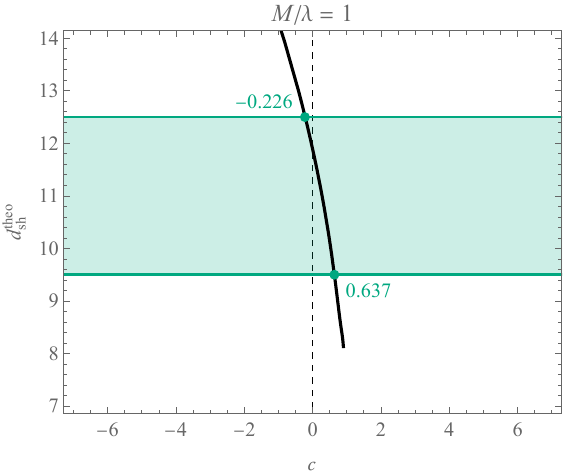} (a)\qquad
    \includegraphics[width=6.5cm]{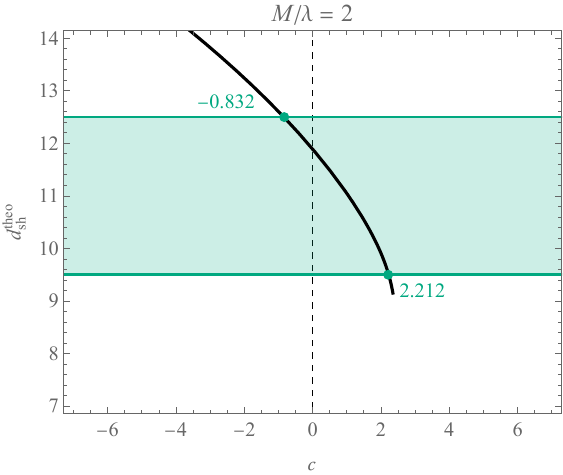} (b)
    \includegraphics[width=6.5cm]{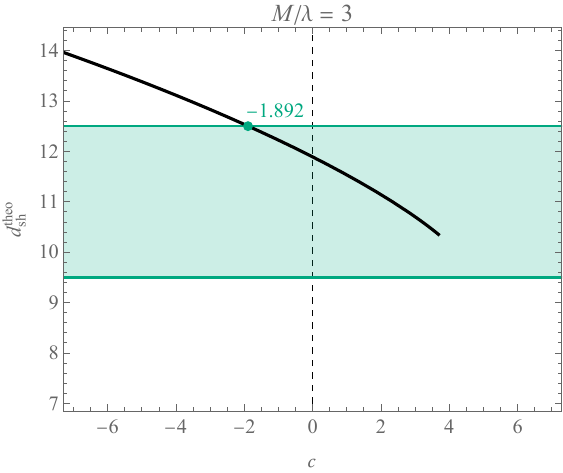} (c)\qquad
    \includegraphics[width=6.5cm]{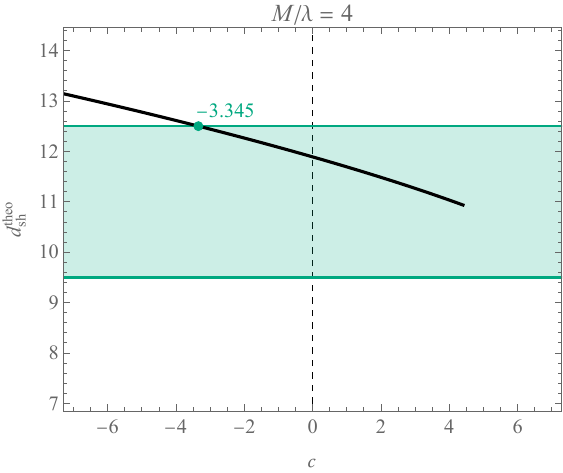} (d)
    \caption{The $\mc$-profile of the theoretical shadow diameter $d_\mathrm{sh}^{\mathbf{theo}}=2b_{\mathrm{sec}}$ of the PHBH for $\vartheta_O=17^\circ$, compared with the observed shadow diameter $d_{\mathrm{sh}}^{\mathrm{M87^*}}$ of M87*, within the $1\sigma$ confidence interval, for four different values of the $M/\lambda$ ratio.
    To plot the curves, we considered a point on the ring corresponding to $r_S=20$ at $\varphi=3\pi/2$.
    The unit of length is taken as the black hole mass $M$.}
    \label{fig:EHTconstraintsM87_new}
\end{figure}

In this subsection, we use the above concepts to derive new constraints on the $\mc$-parameter. To proceed, we calculate the shadow diameter of a black hole using the formula \cite{Bambi:2019tjh}
\begin{equation}
    d_\text{sh} = \frac{r_O \theta_d}{M},    \label{eq:dsh}
\end{equation}
which, based on the astrophysical characteristics of M87* presented earlier, gives $d_{\mathrm{sh}}^{M87^{*}}=11\pm1.5$. Assuming that, by regarding M87* as a PHBH, the secondary image coincides with the photon ring in the GRMHD simulations, we let the theoretical shadow diameter be $d_{\mathrm{sh}}^{\mathrm{theo}}=2b_{\mathrm{sec}}$. By using Eq. \eqref{eq:bphi_1}, we then match the theoretical shadow diameter of the PHBH, $d_{\mathrm{sh}}^{\mathrm{theo}}$, with the observed shadow diameter of M87*, $d_{\mathrm{sh}}^{\mathrm{M87^{*}}}$, for four different values of the $M/\lambda$ ratio, as shown in Fig. \ref{fig:EHTconstraintsM87_new}.
Additionally, in Table \ref{tab:3}, we summarize the constraints on the $\mc$-parameter obtained from this analysis.
\begin{table}[h]
	\centering
	\begin{tabular}{cccccccccccc}
		\toprule
		 &\multicolumn{2}{c}{$M/\lambda=1$}& &\multicolumn{2}{c}{$M/\lambda=2$}& &\multicolumn{2}{c}{$M/\lambda=3$}& &\multicolumn{2}{c}{$M/\lambda=4$}\\
		\cmidrule{2-3} \cmidrule{5-6} \cmidrule{8-9} \cmidrule{11-12}
		
		{} & {upper} & {lower} &  {} & {upper} & {lower}  & {} & {upper} & {lower}&  {} & {upper} & {lower}\\
		\midrule
		$\text{M87*}$ & $-0.2261$ & $0.6374$ & & $-0.8317$ & $2.2119$ & & $-1.8917$ & - & & $-3.3451$ & -\\
		\bottomrule
	\end{tabular}
 \caption{The allowable $\mc$-parameter values, obtained from the curves in Fig. \ref{fig:EHTconstraintsM87_new}, corresponding to the theoretical shadow diameter of the PHBH, by considering $b_{\mathrm{sh}}^{\mathbf{theo}}=2b_{\mathrm{sec}}$, which aligns with the EHT observations of M87* within the $1\sigma$ confidence interval.}
 \label{tab:3}
\end{table}
These constraints were derived based on the assumption that the light ring observed in the GRMHD simulation of M87* corresponds not to the boundary of the real inner black hole shadow, but to the primary image of the accretion disk. Moreover, we considered the furthest possible point from the inner shadow by taking the angular position $\varphi=3\pi/2$ on the $b_{\mathrm{sec}}$ ring (see Fig. \ref{fig:b1_M87images_rS}(b,c)), generating conditions similar to those in the GRMHD simulation of M87*. Therefore, the obtained constraints are novel and, from the perspective of the shadow diameter, more reliable. However, referring to the data summarized in Tables \ref{tab:2} and \ref{tab:3}, the difference between the proposed constraints is significant, and a consistency between such constraints needs to be established. By comparing these constraints for M87*, one can initially rely on the $M/\lambda=1$ case, as it is the most intuitive and natural choice. Secondly, by comparing the upper and lower limits given in these two tables, it is evident that the most reliable choice is to constrain the primary hair as $-0.2261 \leq \mc \leq 0.0553$. This domain is covered by all other $M/\lambda$ ratios and is consistent with both approaches used in the observational tests.

{It is important to note, however, that as indicated in Ref. \cite{erices_thermodynamic_2025}, PHBHs with their scalar hair parameter within the range $-9.0969 \leq \mc \leq 1.3191$ are locally unstable. Therefore, the PHBH models proposed in this section, based on the EHT constraints, are also locally unstable. This local instability highlights the need for considering stabilizing extensions of the theory. In particular, Degenerate Higher-Order Scalar-Tensor (DHOST) frameworks \cite{Langlois:2015cwa, Langlois:2017mdk} provide promising avenues to embed these solutions within a more general and potentially stable scalar-tensor gravity context. Investigations into the thermodynamic and dynamical stability within such extended theories remain an important direction for future work.}

%%%%%%%%%%%%%%%%%%
\subsection{Effects of the primary hair on the higher-order images}

\begin{figure}[htp]
    \centering
    \includegraphics[width=7.5cm]{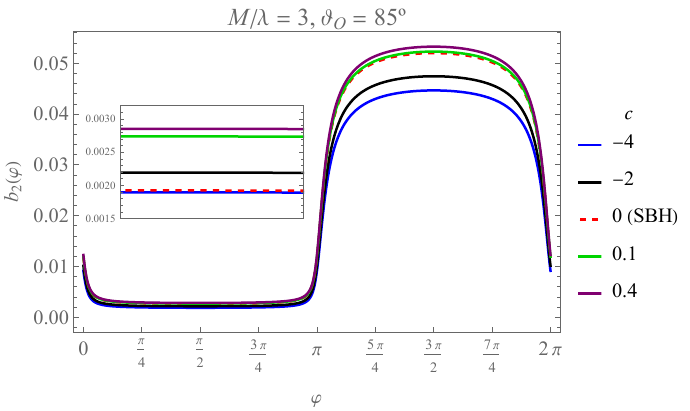} \quad
    \includegraphics[width=7.5cm]{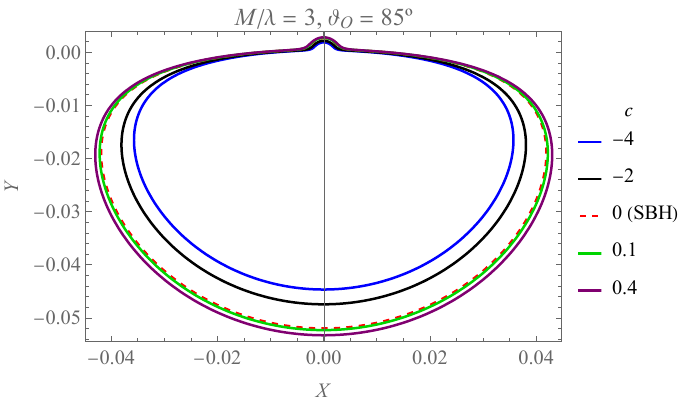} (a)
    \includegraphics[width=7.5cm]{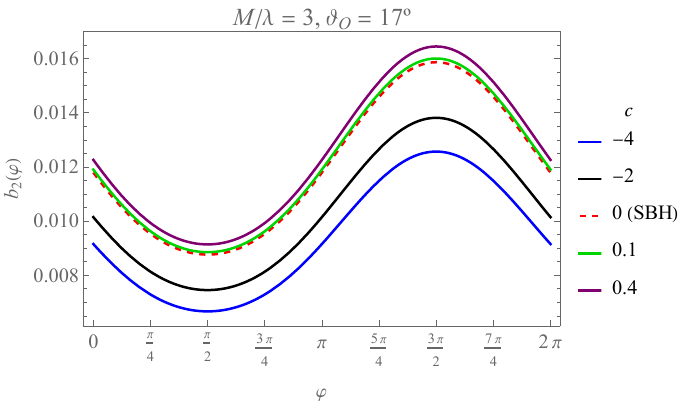} \quad
    \includegraphics[width=7.5cm]{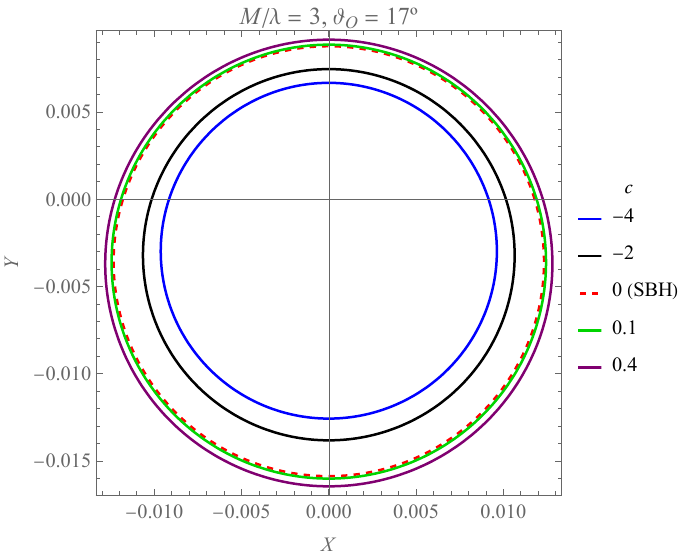} (b)
    \caption{The behavior of the tertiary image considering M87* as a PHBH, plotted for $M/\lambda = 3$, two different cases of $\vartheta_O = 85^\circ, 17^\circ$, and several values of the $\mc$-parameter. The left panels show the $\varphi$-profiles of $b_2(\varphi)$, while the right panels show the deviation function $\tilde{b}_2(\varphi)$. In these diagrams, we have assumed $r_S = r_c$.}
    \label{fig:M_M87_b2_c}
\end{figure}

As the image order increases, the deviation of the photon rings from circularity becomes progressively smaller, starting with the $b_2$ image, which almost coincides with $b_\infty = b_c$, the actual theoretical shadow size of any black hole. This deviation can be calculated as \cite{tsupko_shape_2022}
\begin{equation}
\tilde{b}_n(\varphi) = b_n(\varphi) - b_c.
    \label{eq:btilde_phi}
\end{equation}
To demonstrate the effects of the primary scalar hair on the higher-order images of the PHBH, we consider the tertiary image of M87* assuming it to be a PHBH in Fig. \ref{fig:M_M87_b2_c}, for different values of the $\mc$-parameter, with $M/\lambda$ fixed. 
In this figure, we show the changes in the $b_2$ images versus the polar angle $\varphi$, for different $\mc$-parameter values. It can be observed that as this parameter increases from negative to positive values, the radius of the tertiary image increases, while the deviations from circularity become less pronounced. This is particularly evident from the plots of the $\tilde{b}_2$ function on the observer's screen.

\begin{figure}[htp]
    \centering
    \includegraphics[width=7.5cm]{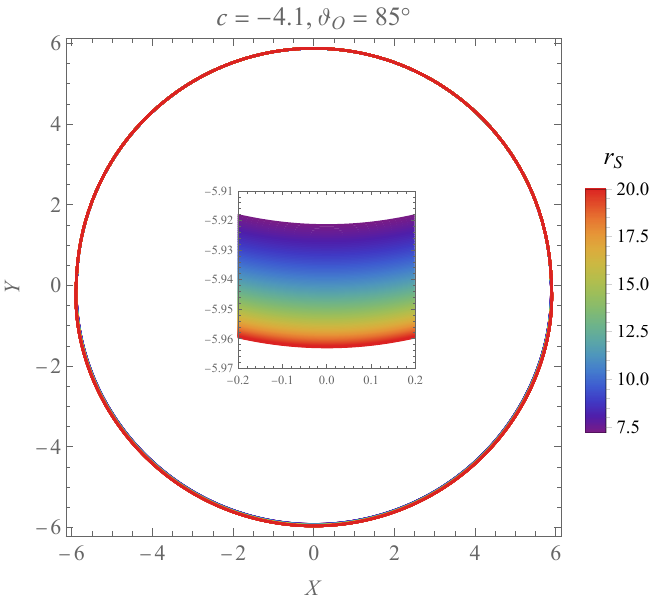} (a)
    \includegraphics[width=7.5cm]{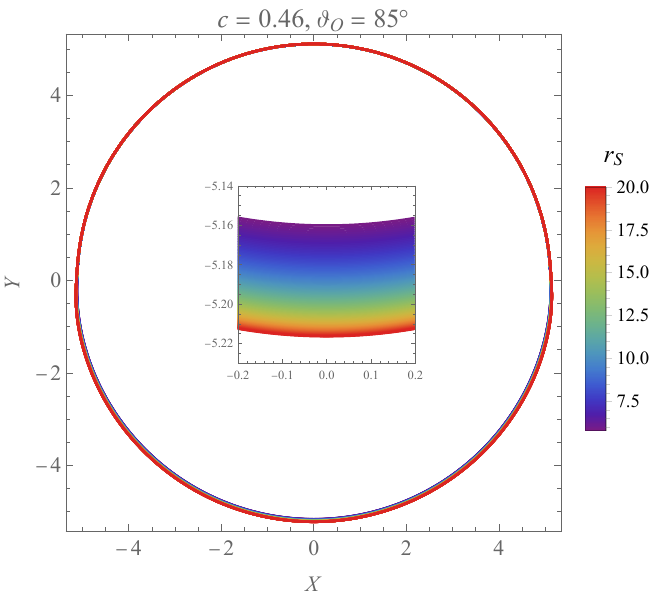} (b)
    \caption{The behavior of the tertiary image $b_2$, assuming M87* to be a PHBH, plotted for $M/\lambda = 3$ and $\vartheta_O = 85^\circ$, with two different scenarios for the $\mc$-parameter, and $r_S$ ranging from $r = r_c$ to $r = 20$.
    The radius of the inner black hole shadow is (a) $b_c = 5.877$, and (b) $b_c = 5.107$.
    The unit of length along the axes is taken as the black hole mass $M$.}
    \label{fig:M_M87_b2_rS}
\end{figure}

\begin{figure}[htp]
    \centering
    \includegraphics[width=7.5cm]{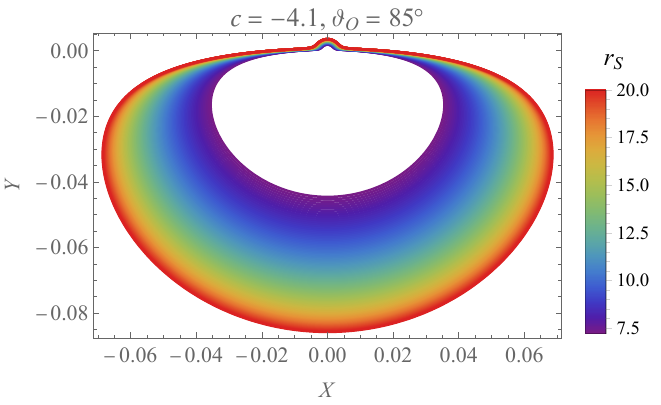} (a)
    \includegraphics[width=7.5cm]{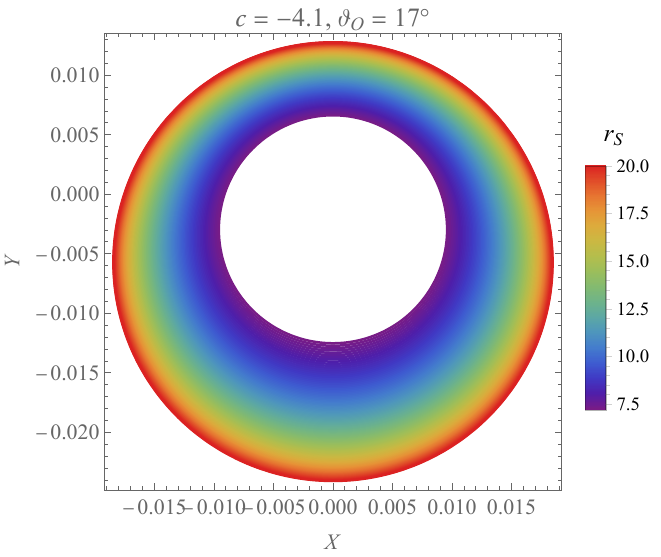} (b)
    \includegraphics[width=7.5cm]{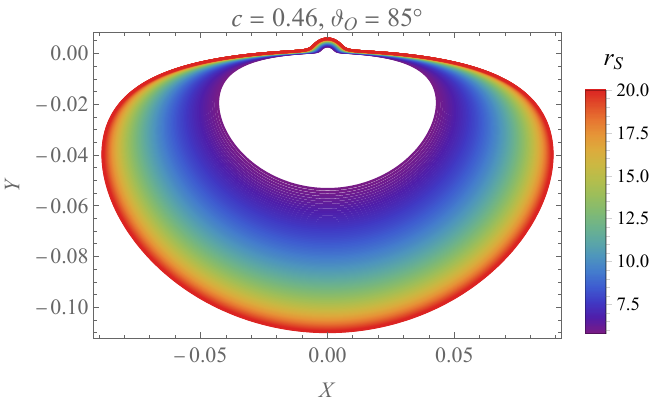} (c)
    \includegraphics[width=7.5cm]{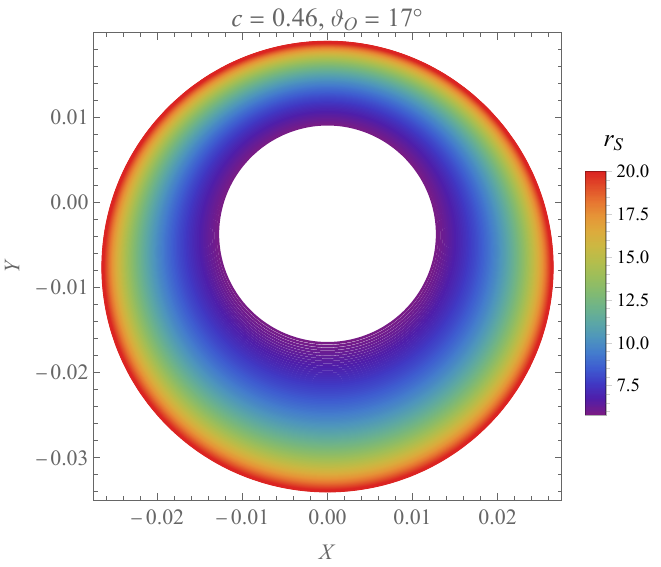} (d)
    \caption{The behavior of the deviation function $\tilde{b}_2$, assuming M87* to be a PHBH, plotted for $M/\lambda = 3$ and the two inclinations $\vartheta_O = 85^\circ, 17^\circ$, with two different scenarios for the $\mc$-parameter, and $r_S$ ranging from $r = r_c$ to $r = 20$.
    The unit of length along the axes is taken as the black hole mass $M$.}
    \label{fig:M_M87_bt2_rS}
\end{figure}

However, it should be noted that as the deviations from circularity are small, the $b_2$ images are very close to the actual black hole shadow, making them thin and demagnified. As a result, they are of lesser importance in GRMHD simulations of black hole shadows, as they are not easily observable. This is further demonstrated in Fig. \ref{fig:M_M87_b2_rS}, where we show the behavior of the $b_2$ images at high inclination.
As seen in the diagrams, even at large inclinations, a thin circular-shaped image is still observable, close to the inner shadow of the black hole. The thickest part of the images appears around $\varphi \approx 3\pi/2$, occupying a small range on the $Y$-axis of the observer's screen. Additionally, to further highlight how close the images are to circularity, Fig. \ref{fig:M_M87_bt2_rS} depicts the behavior of the deviation function $\tilde{b}_2$ for the same model chosen in Fig. \ref{fig:M_M87_b2_rS}, for two different inclination angles. 

{Thus, the tertiary image appears as a very thin structure, located very close to the inner black hole shadow, making its separation challenging in practical observations. Moreover, current VLBI angular resolutions, flux limitations, and $(u,v)$ coverage further hinder the detectability of such higher-order images, as they are effectively unresolved and faint in GRMHD simulations. As discussed in Ref.~\cite{ayzenberg_testing_2022}, these observational constraints imply that higher-order images (black hole shadow subrings) are unlikely to provide significant astrophysical information with current instrumentation, analogous to the challenges faced in resolving black hole subrings.}

%%%%%%%%%%%%%%%%%%%%%
\section{Summary and conclusions}\label{sec:conclusions}

In this work, we have investigated the gravitational lensing effects of a static, asymptotically flat black hole with primary scalar hair within Beyond Horndeski gravity, focusing on the strong lensing regime. Our study extends previous constraints on the scalar hair parameter obtained from thermodynamic stability and black hole shadow considerations by incorporating lensing observables derived from the EHT data for M87* and Sgr A*. Through a comprehensive analysis of photon trajectories and their corresponding lensed images, we have established new, more precise constraints on the scalar hair parameter, offering additional insights into the nature of black holes in modified gravity theories. 

A key aspect of our study was the formulation of the deflection angle in the strong lensing regime. Utilizing the Virbhadra-Ellis lens equation and Bozza’s analytical approach, we demonstrated how the primary scalar hair parameter influences the bending of light rays around the black hole. Our results show that the presence of primary scalar hair alters the location and properties of the photon sphere, leading to significant modifications in the lensing observables. Notably, we found that black holes with negative scalar hair parameter exhibit a larger deflection angle compared to their general relativistic counterparts, while those with positive parameter display reduced deflection effects. This observation directly impacts the formation of relativistic images and the RERs observed in black hole imaging.

Furthermore, we examined the angular separation and relative magnification of the strong lensing images, both of which provide astrophysical constraints on the black hole’s parameters. By applying our theoretical framework to the EHT observations of M87* and Sgr A*, we derived constraints on the scalar hair parameter within the 1$\sigma$ confidence level. Our findings indicate that for M87*, the allowed range for the scalar hair parameter is $-0.487 \leq c \leq 0.055$, while for Sgr A*, the constraints are slightly different, falling within the interval $-0.301 \leq c \leq 0.491$. These constraints are more stringent than those obtained from shadow analysis alone, as strong lensing effects provide an independent observational handle on the black hole metric.

Additionally, we explored the impact of primary scalar hair on the lensed images of a thin accretion disk. By numerically computing the shapes and positions of different-order images, we provided an alternative method for constraining the black hole parameters. Our results suggest that the observed black hole shadow in EHT images corresponds to the secondary image of the emitting disk rather than the actual shadow. This interpretation allows for a novel approach to parameter estimation, wherein the angular diameter of the emission ring is directly related to the black hole’s lensing properties. In this context, we found that the constraints derived from the accretion disk images are consistent with those obtained from strong lensing analysis, further reinforcing the robustness of our findings.

We also discussed the role of higher-order images in black hole lensing and their implications for astrophysical observations. Our results show that while these images encode important information about the underlying spacetime geometry, their faintness and proximity to the primary shadow limit their observational significance. Consequently, strong lensing constraints derived from the brightest relativistic images remain more reliable.
\raggedbottom

Finally, an important conclusion of our study is the local instability of black hole solutions satisfying the derived constraints. As previously noted, models with primary scalar hair within the range $-9.097 \leq c \leq 1.319$ exhibit instabilities, raising concerns about their astrophysical relevance. This suggests that while observational constraints favor specific values of the scalar hair parameter, the physical viability of these models requires further investigation, particularly in the context of dynamical stability and perturbative analyses.
\raggedbottom

{In summary, this work provides a comprehensive analysis of strong gravitational lensing in Beyond Horndeski gravity, deriving new observational constraints on the primary scalar hair parameter using the EHT data. These results advance the testing and refinement of alternative gravity theories through astrophysical measurements, offering valuable insights into black hole properties beyond general relativity. Future research may build upon this foundation by investigating strong lensing effects in a wider range of black hole solutions with scalar hair recently explored in the literature~\cite{PhysRevD.109.064024,PhysRevD.110.024044}. Although such studies pose significant computational challenges, they hold promise for deepening our understanding of the physical viability and thermodynamic stability of these extended scalar-tensor models~\cite{PhysRevD.110.L101502,2025arXiv250502368M}.}

{Finally, we note that the static black hole solutions studied here, as well as similar configurations, can in principle be extended to rotating counterparts via the Newman–Janis algorithm~\cite{Newman:1965}. The strong gravitational lensing analysis could then be applied to these rotating geometries using the formalism developed in Ref.~\cite{bozza_strong_2007}. Such an extension would offer a more realistic framework for astrophysical applications and is left for future work.}

%%%%%%%%%%%Acknowledgements 
\section*{Acknowledgements}
This work is supported by Universidad Central de Chile through the project No. PDUCEN20240008.

%%%%%%%%%%%%%%%%appendices
%\appendix

%%%%%%%%%%%%%%

%%%%%%%%%%%%%References
\bibliographystyle{ieeetr}
\bibliography{biblio.bib}

\end{document}